\pdfoutput=1

\documentclass[11pt]{article}

\usepackage[preprint]{acl}

\usepackage{times}
\usepackage{latexsym}

\usepackage[T1]{fontenc}

\usepackage[utf8]{inputenc}

\usepackage{microtype}

\usepackage{inconsolata}

\usepackage{graphicx}

\author{Kartik Chandra (MIT)\And Katherine Collins (Cambridge) \And Will Crichton (Brown) \AND Tony Chen (MIT) \And Rachit Nigam (Cornell) \And Adrian Weller (Cambridge) \AND Tzu-Mao Li (UCSD) \And Joshua B. Tenenbaum \And Jonathan Ragan-Kelley (MIT)}

\newcommand{\watchat}[0]{{WatChat}}
\newcommand{\watchatjs}{{WatChat@JS}}
\newcommand{\watchatgit}{WatChat@Git}

\title{
\watchat: Explaining perplexing programs by debugging mental models
}

\usepackage{xcolor}
\usepackage{graphicx}
\usepackage{amsmath}
\usepackage{stmaryrd}
\usepackage{centernot}
\usepackage{mdframed}
\usepackage{soul}
\mdfsetup{skipabove=5pt}

\usepackage[scaled]{beramono}
\usepackage[T1]{fontenc}

\usepackage{booktabs}
\usepackage{listings}
\lstdefinelanguage{rosette}{
    morekeywords={define, cond, and},
    sensitive=false,
    morecomment=[l]{;},
    morestring=[b]"
}

\begin{document}
\maketitle
\begin{abstract}
Often, a good explanation for a program's unexpected behavior is a bug in the programmer's code. But sometimes, an even \emph{better} explanation is a bug in the programmer's \emph{mental model} of the language or API they are using. Instead of merely debugging our current code (``giving the programmer a fish''), what if our tools could directly debug our mental models (``teaching the programmer to fish'')?
In this paper, we apply recent ideas from computational cognitive science to offer a principled framework for doing exactly that. Given a ``why?'' question about a program, we automatically infer potential misconceptions about the language/API that might cause the user to be surprised by the program's behavior---and then analyze those misconceptions to provide explanations of the program's behavior. Our key idea is to formally represent misconceptions as counterfactual (erroneous) semantics for the language/API, which can be inferred and debugged using program synthesis techniques.
We demonstrate our framework, \watchat, by building systems for explanation in two domains: JavaScript type coercion, and the Git version control system. We evaluate \watchatjs\ and \watchatgit\ by comparing their outputs to experimentally-collected human-written explanations in these two domains: we show that \watchat's explanations exhibit key features of human-written explanation, unlike those of a state-of-the-art language model.
\end{abstract}

\begin{figure*}
    \includegraphics[width=\linewidth,page=1,trim={0 12cm 0 0},clip]{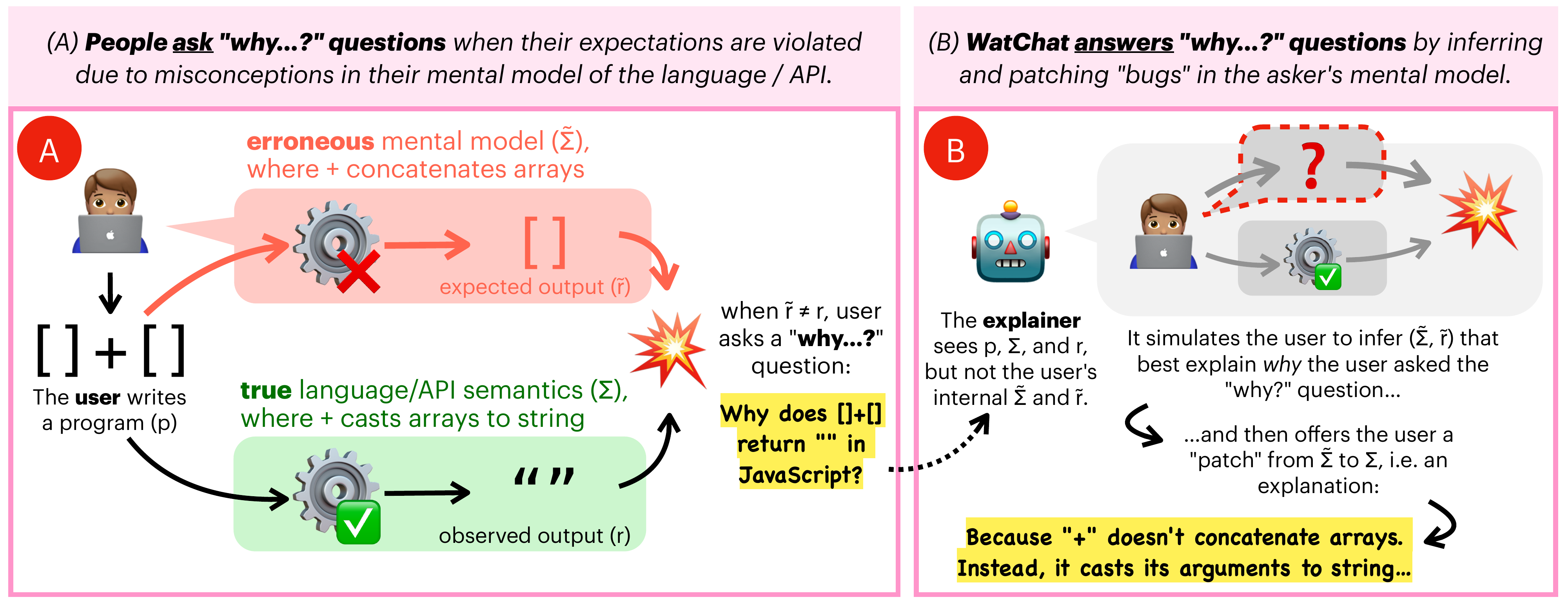}
\caption{Overview of the \watchat\ framework for explanation---here, applied to explain the output of a JavaScript program. Panel ``A'' shows \watchat's model of how programmers come to ask ``why?'' questions. Panel ``B'' shows how \watchat\ reasons over the model in Panel ``A'' to find a good explanation.}\label{fig:teaser}
\end{figure*}

\section{Introduction}
\renewcommand{\thefootnote}{\relax} 
\footnotetext{This is a preprint of work presented in early-stage non-archival form at the ACL 
Natural Language Reasoning and Structured Explanations Workshop.}
\renewcommand{\thefootnote}{\arabic{footnote}}
In a well-known 2012 tech talk called ``Wat'' \citep{wat2012bernhardt},
Gary Bernhardt asks the audience, ``Does anyone know, in JavaScript, what array plus array (\lstinline{[] + []}) is? Is it empty array? Type error?'' When he types this expression into a REPL, the audience bursts out laughing to see that it actually returns \lstinline{""}, the empty string. As the audience's reaction shows, such behavior is surprising even to expert programmers: it prompts them to exclaim ``WAT!'' --- and then to ask \emph{``why?''}.

%
A typical explanation of this behavior, e.g.\ on a website like Stack~Overflow \citep{stackoverflow2012wat}, gives two key pieces of information: First, when given non-numbers, JavaScript's \lstinline{+} operator converts its operands to strings for concatenation. Second, Arrays cast to strings by joining their elements with commas (without outer square brackets), so the empty array casts to the empty string.

Why do these two statements together form a satisfying explanation of the fact that \lstinline{[] + []} gives \lstinline{""}? If that seems like an odd question to ask, note that there are many other possible responses that could theoretically have passed for ``explanations'' of this behavior. For example, ``because the ECMAScript specification says so'' would technically be a correct explanation, as would a tedious step-by-step trace of how the given program is parsed, JIT-compiled, and executed on hardware. But of course, intuitively we find such explanations to be dissatisfying, uninsightful, unhelpful---and in some cases, even condescending.

It is this intuition that we are after in this paper. \textbf{What makes a good explanation?} And how do humans recognize and construct good explanations for one another? These are important questions that lie squarely at the intersection of programming languages, software engineering and human-computer interaction: we would like our tools to be able to explain themselves to us in a helpful, human-like way. But even beyond these fields, the study of explanation has important consequences for disciplines ranging from education \citep{lombrozo2006structure, anderson1995tutor} to artificial intelligence \citep{miller2019explanation} to scientific understanding and progress \citep{strevens2011depth}.

In this paper, we take inspiration from empirical cognitive science work on explanation (reviewed in Section~\ref{sec:background}) to design systems that give helpful, human-like explanations of unexpected program behavior. Our guiding premise is that explanation is a cooperative social interaction: by asking a ``why?'' question, the asker signals that they do not understand an observed phenomenon, and requests a correction or addition to their mental model (Figure~\ref{fig:teaser}A). The explainer in turn identifies possible gaps in the asker's mental model that would cause them to ask ``why?,'' and makes a relevant ``because'' statement that helpfully fills those gaps in (Figure~\ref{fig:teaser}B).


Putting these ideas into practice, we design a framework for giving good, human-like responses to ``why?'' questions. Our key idea is that we can formally represent the user's erroneous mental model as \emph{counterfactual (erroneous) semantics} for the language or API they are using. The user expects their program to produce a result under these counterfactual semantics, and is surprised when they see that the true result (produced under the true semantics) is different from what they expect. The task of explanation then becomes to \emph{infer} and \emph{debug} these counterfactual semantics, i.e.\ to correct the user's mental model. We call this framework \watchat, and we formalize it in Section~\ref{sec:watchat}.

The \watchat\ framework can be instantiated to give explanations in a variety of domains. For example, in Section~\ref{sec:watchatjs} we show how to apply \watchat\ to the domain of JavaScript type coercion to produce \watchatjs, a system that can explain the surprising behaviors from the ``Wat'' talk (as well as a variety of other cases). If you ask \watchatjs\ ``Why does \lstinline{[] + []} return \lstinline{""}?'' then \watchatjs\ does what Bernhardt does in his talk: it asks you if you expected the empty array as output. If you say yes, then \watchat\ explains that {\color{gray}``The \lstinline{+} operator does not concatenate arrays; instead, it casts the arrays to strings and concatenates the resulting strings.''} If you then go on to ask, ``But then why isn't the output \lstinline[]$"[][]"$?'' \watchat\ responds, {\color{gray}``When coerced to strings, arrays don't have \lstinline[]$[]$s around them. Hence, \lstinline[]$[]$ gives \lstinline{""} and thus \lstinline[]$[] + []$ gives \lstinline{""}.''}
Notice that this explanation closely matches the one found on Stack~Overflow, offering the same two pieces of information: first, about \lstinline{+} coercing arrays to string, and second, about the missing square brackets.

Of course, the \watchat\ framework is not limited to JavaScript. For example, as we show in Section~\ref{sec:watchatgit}, \watchat\ can also be applied to give explanations about the Git version control system. The resulting system, \watchatgit, can explain a variety of surprises that a programmer might encounter while using Git. For example, it can explain why a user got an error when trying to open a file after removing it from the repository with \lstinline{git rm}. It does not explain this by simply declaring that ``no such file exists,'' as an error message from \lstinline{cat} might. Nor does it do so by saying ``because you just deleted the file with \lstinline{git rm}.'' Rather, \watchatgit\ directly addresses the fundamental conceptual issue at play, which is that the \lstinline{git rm} command not only stages the file's removal from the current branch, but also actually deletes the file from disk (unless the \lstinline{--cached} flag is passed in).

%
%
%
%
In Section~\ref{sec:eval}, we evaluate these two systems by comparing their outputs to experimentally-collected human-written explanations, as well as to explanations generated by a state-of-the-art language model. Our evaluation shows how \watchat\ captures several key properties of good, human-like explanations.
Finally, in Section~\ref{sec:discussion} we discuss the limitations of our system, and conclude with scope for future work.

Our implementations of \watchatjs\ and \watchatgit\ are available at [GitHub links redacted for anonymity]. 

\begin{figure}
\centering
\includegraphics[width=\linewidth,page=3,trim={20cm 24cm 20cm 0},clip]{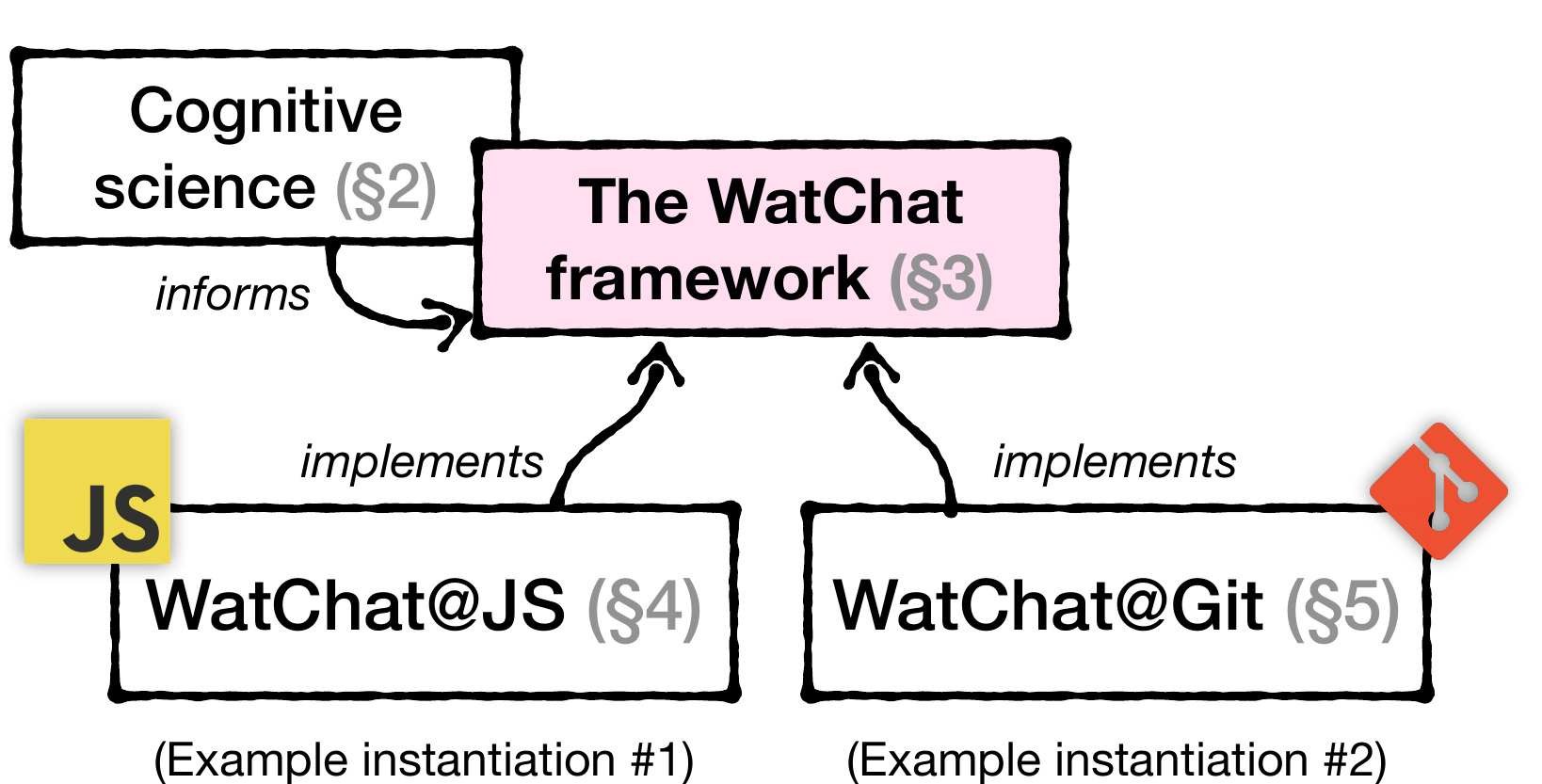}
\caption{High-level overview of this paper. We describe \watchat, a general framework for explanation, and we apply it to produce two systems: \watchatjs\ and \watchatgit.}\label{fig:overview}
\end{figure}

\section{The \watchat\ framework for explanation}\label{sec:watchat}

WatChat is a conceptual framework for building systems that explain the behavior of programs to a person based on that person's erroneous mental model of the programming language or API they are using. Building on \citeauthor{chandra2024explanation}'s work, the key idea behind \watchat\ is to infer possible misconceptions the user has based on the program being asked about, as well as (if provided) the user's expectation. This is a technical challenge: we must not only formalize the vast space of erroneous mental models a user might have, but also efficiently search through it.
Our solution is to model users' mental models as \emph{counterfactual (erroneous) semantics} for the language or API they are using (recall Figure~\ref{fig:teaser}A). Given a program to be explained, we use program synthesis techniques \citep{gulwani2017program} to synthesize counterfactual semantics under which the given program gives a different result from the standard language/API semantics. (This is a somewhat unusual application of program synthesis: our goal is not to synthesize the program, as is traditional, but rather the \emph{semantics} under which the given program behaves as the user expects. Conveniently, we can still apply program synthesis techniques because semantics can be represented by interpreters, which are themselves programs that can be synthesized.)
Once the counterfactual semantics have been inferred, then explanation takes the form of patching bugs in it, i.e.\  debugging the user's mental model (recall Figure~\ref{fig:teaser}B).

In the rest of this section, we will give a formal description of the \watchat\ framework.
%
%
\watchat\ takes as input a program $p$ whose behavior is to be explained to the user. Under the true semantics of the language or API, which we will denote with  $\Sigma$, the program $p$ produces some result $r$. We will write $\llbracket p \rrbracket_\Sigma = r$ to mean ``$p$ produces result $r$ under the semantics $\Sigma$.''

Now, because the user is asking for an explanation of $p$'s behavior, we assume that they instead expected some alternate result $\widetilde{r} \not= r$. This alternate expectation must have been caused by the user having one or more misconceptions about the language or API's semantics in their mental model. A good, helpful explanation of $p$'s behavior should infer these misconceptions and correct them.

To operationalize this idea, we model the user's misconceptions as counterfactual, erroneous semantics $\widetilde{\Sigma}$ for the language, under which $p$ actually does produce $\widetilde{r}$, i.e.\ $\llbracket p \rrbracket_{\widetilde{\Sigma}} = \widetilde{r}$. Our goal is thus to infer $\widetilde{\Sigma}$ given $p$. (Note that we do not necessarily know $\widetilde{r}$ in advance --- depending on the ``why?'' question asked, a user may or may not have revealed their expectation. As we will see below, our system handles ambiguity about $\widetilde{r}$ by asking clarification questions when needed.)
We perform this inference task with an unconventional application of program synthesis: we synthesize a ``misinterpreter'' for $\widetilde{\Sigma}$ that satisfies the constraint that $\llbracket p\rrbracket_\Sigma \not= \llbracket p\rrbracket_{\widetilde{\Sigma}}$ (we borrow the term ``misinterpreter'' from \citet{smol2023}). This application of synthesis is unconventional in the sense that while synthesis traditionally searches for a program with respect to a fixed interpreter, here we do the opposite: search for an \emph{interpreter} with respect to a fixed \emph{program}.

Of course, this is a severely under-determined problem. Given only $p$, there may be infinitely many pairs $\langle \widetilde{\Sigma}, \widetilde{r} \rangle$ that satisfy our criteria: for example,  for each $\widetilde{r} \not= r$, consider the degenerate semantics $\widetilde{\Sigma}_{\widetilde{r}}$ where \emph{all} programs vacuously output $\widetilde{r}$. Intuitively, we know that these are unlikely mental models for the user to have---indeed, the set of \emph{plausible} mental models is extremely sparse in the set of \emph{possible} mental models.
Thus, we reason probabilistically: we define a probability distribution over misinterpreters based on our prior belief about how likely various misconceptions are. Then, we infer $\widetilde{\Sigma}$ and $\widetilde{r}$ on a maximum a posteriori (MAP) basis, conditioned on the observation that $\widetilde{r} \not= r$. If there are multiple equally-good explanations, we ask the user for clarification about their expectations (e.g.\  by asking ``Did you expect $\widetilde{r_1}$ or $\widetilde{r_2}$?''). The strong inductive bias given by the prior distribution, combined with the ability to ask clarification questions, allows us to find reasonable solutions to this underconstrained problem.\footnote{Readers might wonder at this point if this problem is tractable. It turns out that program synthesis techniques can be applied to perform this search efficiently. For example, the implementation in Section~\ref{sec:watchatjs} is able to search through a space of $2^{32}$ potential mental models within milliseconds using a powerful constraint solver.}

Finally, once the user's erroneous mental model $\widetilde{\Sigma}$ has been inferred, it can be used to produce a concise explanation of the program's behavior. Rather than outputting a full causal trace of the program's execution, we \emph{selectively} output only the relevant portions of the trace, i.e.\ those that would be affected by misconceptions in $\widetilde{\Sigma}$. Each relevant step of the trace is accompanied with a message correcting the corresponding misconception.

Notice that this framework for explanation fulfills all of \citeauthor{miller2019explanation}'s desiderata: it is contrastive (it explains why $p$ did \emph{not} produce $\widetilde{r}$), it is selective (it only communicates the relevant misconceptions in $\widetilde{\Sigma}$, not the full description of $\Sigma$ or the full evaluation trace of $p$), it is causal (the misconceptions are causally relevant to producing $r$), and it is social (it makes inferences about the asker's mental state, and engages in interactive clarification when necessary).

We instantiate the \watchat\ framework in two domains: JavaScript and Git. These two domains demonstrate the flexibility of \watchat: our framework can be applied to programming languages as well as developer tools, and furthermore it is compatible with a variety of implementation strategies and search algorithms. For example, \watchatjs\ was implemented by writing a model of JavaScript's semantics and using a constraint solver, whereas \watchatgit\ was implemented by patching an existing Git client and using parallelized enumerative search. Table~\ref{tab:comparison} compares the implementations of \watchatjs\ and \watchatgit. Complete details, with worked examples, are provided in the Appendix.

\section{Evaluation}\label{sec:eval}

We set out to understand human intuitions behind explanation, and to design systems that capture the key features of human explanation. In this section, we evaluate \watchatjs\ and \watchatgit\ to see whether they do indeed align with human intuitions about explanation.

At a high level, our evaluation compares explanations generated by \watchatjs\ and \watchatgit\ to explanations generated by humans, and those generated by a state-of-the-art large language model (GPT-4). Our goal is to see how these explanations are similar and how they differ, through the lens of \citeauthor{miller2019explanation}'s desiderata (Section~\ref{sec:background}). In Section~\ref{sec:eval-methods}, we describe our methodology: how we selected the scenarios to generate explanations for, details of how we elicited explanations from humans and GPT-4, and how we coded explanations for comparison. Then, in Section~\ref{sec:eval-analysis}, we analyze the coded explanations. Appendix Section~\ref{sec:eval-ttv} outlines threats to validity of this evaluation.

It is important to note that we have \emph{not} yet tested whether \watchatjs's and \watchatgit's explanations are effective in resolving users' confusions. This is of course a critical feature of explanations, and one we plan to test in the future (see Section~\ref{sec:eval-ttv}). However, it is not the scope of our present work, which is focused specifically on capturing human \emph{explainers'} intuitions about what should be included in an explanation.

\begin{figure*}
    \centering
    \includegraphics[width=\linewidth]{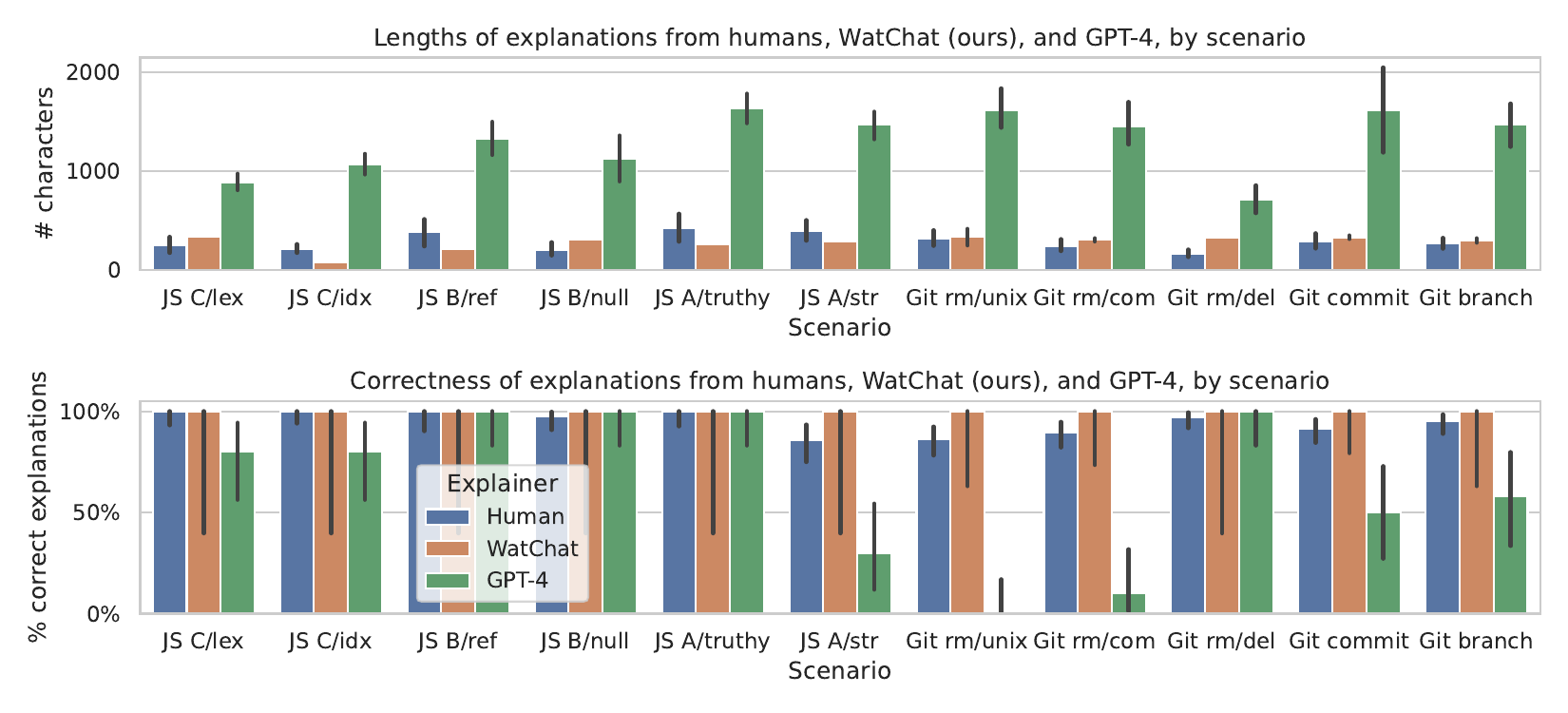}
    \caption{Average length and correctness of explanations across our 11 scenarios. Explanations from \watchat\ and humans are consistently correct and succinct, typically well under 500 characters. Explanations from GPT-4, in contrast, are often incorrect (especially in Git scenarios), and are typically around 1,000--1,500 characters long. All error bars in this and subsequent figures are 95\% confidence intervals. In the bottom graph, error bars are narrower for human and GPT-4 responses than for \watchat's responses: this is because for humans and GPT-4, data is aggregated across several independent responses/rollouts in addition to across the four coders.}\label{fig:summary}
\end{figure*}

\begin{figure*}
    \centering
    \includegraphics[width=\linewidth]{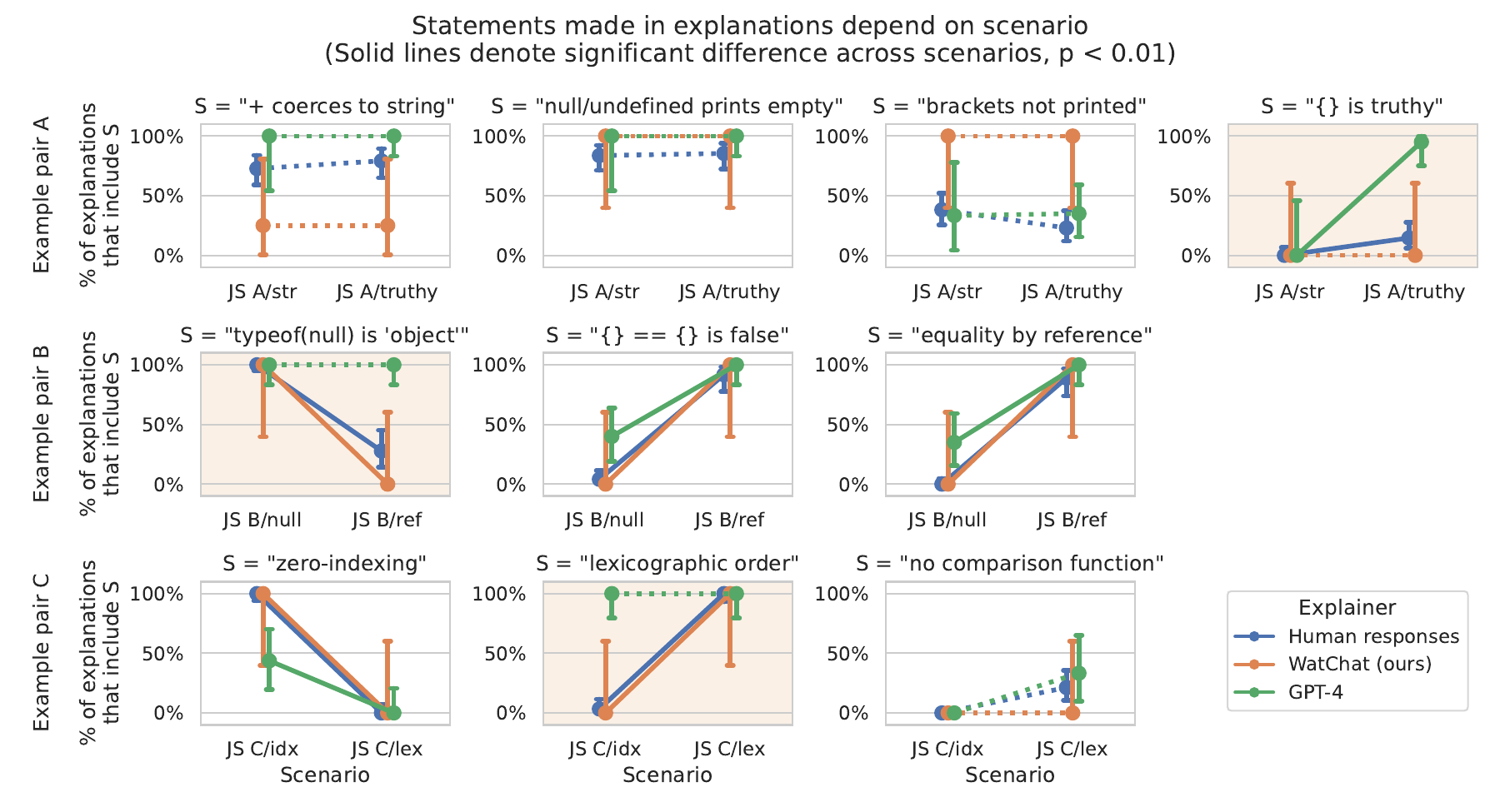}
    \caption{How do explanations depend on the scenario being asked about (Section~\ref{sec:js-analysis})? Here, each row considers one pair of example JavaScript scenarios. Each plot within a row shows how a particular statement $S$'s presence in an explanation varies by scenario. \watchat\ tracks trends in human responses well, but GPT-4 often over-explains. This is particularly salient in the three shaded plots, where GPT-4 (green line) departs from the trend in human (blue) and \watchat\ (orange) explanations. (See note on error bars in Figure~\ref{fig:summary}.)}\label{fig:js-summary}
\end{figure*}

\subsection{Methods}\label{sec:eval-methods}

\paragraph{Choice of scenarios} We collected explanations for several scenarios in both the JavaScript and Git domains (see Appendix). The two domains have different characteristics, which create opportunities for studying different aspects of explanation.

The JavaScript domain has a very rich space of possible expectations (i.e.\ all JavaScript values), and misconceptions in the user's mental model affect their expectations in interesting, fine-grained ways. Thus, we used JavaScript scenarios to study the ways in which the user's stated expectations influence the explanations that humans, \watchatjs, and GPT-4 produce for a given program.


For the JavaScript domain, we collected explanations for the programs listed in Appendix Section~\ref{sec:js-examples}. We chose these scenarios because they came in three natural pairs that show how explanation depends on context and expectation:
\begin{itemize}
    \item The program C/idx with expectation \lstinline{1}, and the program C/lex with expectation \lstinline{4}: these are very similar programs for which \watchatjs\ gives very different explanations.
    \item The program B/null (same as B/ref), with the two different expectations mentioned in the text: these are \emph{identical} programs for which \watchatjs\ very different explanations based on expectation.
    \item The program A/str with expectation \lstinline{"Answer:[true,null][false]"}, and the program A/truthy with expectation \lstinline{"[2,undefined,2]"}: these are different programs for which \watchatjs\ gives the \emph{same} explanation for these expectations (in particular, \emph{without} mentioning ``truthiness'' in the latter case).
\end{itemize}
For the Git domain, we collected explanations for all five scenarios listed in Appendix Section~\ref{sec:git-examples}. We chose these scenarios because they were thematically linked --- this lowered the burden on human participants, because they only had to deeply understand one set of circumstances and only look up the details of a few Git commands. Note that the Git domain has a comparatively smaller space of expectations (mostly related to unexpected errors). However, as we discuss in Appendix Section~\ref{sec:git-examples}, this domain has the interesting property that a rich space of possible erroneous mental models could all lead to the same surprising behavior. Hence, we used Git scenarios to study cases where there are multiple potential erroneous mental models that could account for the user's expectation violation.

\paragraph{Comparison to human-written explanations}
We recruited $N=25$ participants for an on-line experiment, which was conducted with IRB approval. Participants were recruited via university computer science department mailing lists (undergraduate- and graduate-level), as well as from programming-related online message boards like Reddit. Participants were told that they would be shown a series of emails from colleagues asking questions about unexpected behavior in JavaScript and Git. Each email would be from a different colleague of unknown seniority. Participants were instructed to respond to the emails with explanations of the unexpected behavior. They were not required to fix or improve the code. See Appendix~\ref{sec:additional-human-exp} for further details and Appendix~\ref{sec:prompts} for the full text of the instructions. 

\paragraph{Comparison to large language models (LLMs)} We additionally collected explanations from a state-of-the-art language model, which at the time of writing is {OpenAI's} GPT-4~\citep{openai2024gpt4}. Using OpenAI's API, we prompted GPT-4 (specifically, ``gpt-4-1106-preview'') with the instructions given to participants in the study described above (see Appendix~\ref{sec:prompts} for the full prompt, and a discussion of how we adapted the human instructions for the prompt). For each scenario, we independently sampled 5 explanations at the default temperature of 1. Hence, we collected a total of $11\times5=55$ different explanations from GPT-4.

\paragraph{Coding responses} Four authors of this paper independently coded the explanations we collected from humans, \watchat, and GPT-4, based on (a)~whether or not the explanation was correct (i.e.\ did not make any false claims), and (b)~if it was correct, then which \emph{statements} the explanation made. Additional details on coding are included in Appendix~\ref{sec:coding}.

\subsection{Results}\label{sec:eval-analysis}

\begin{figure*}[t]
    \centering
    \includegraphics[width=\linewidth]{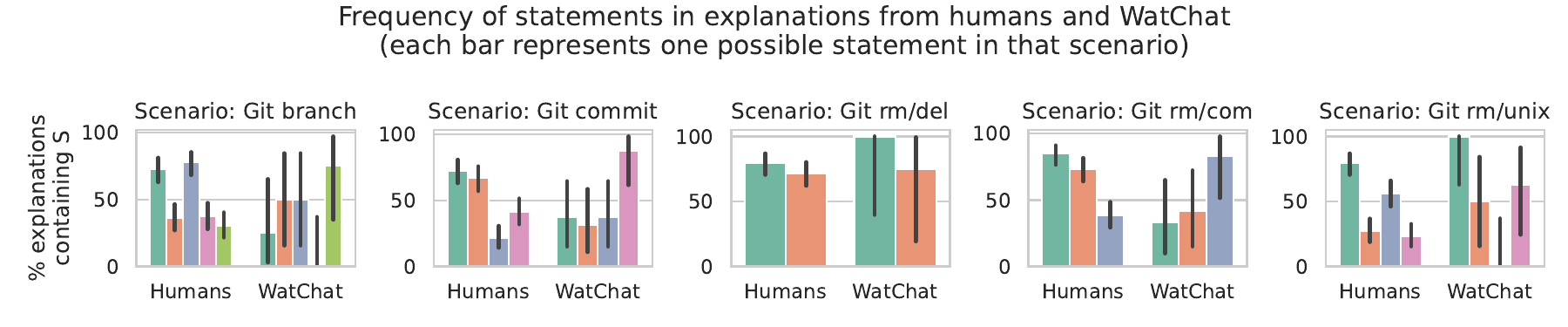}
    \includegraphics[width=\linewidth]{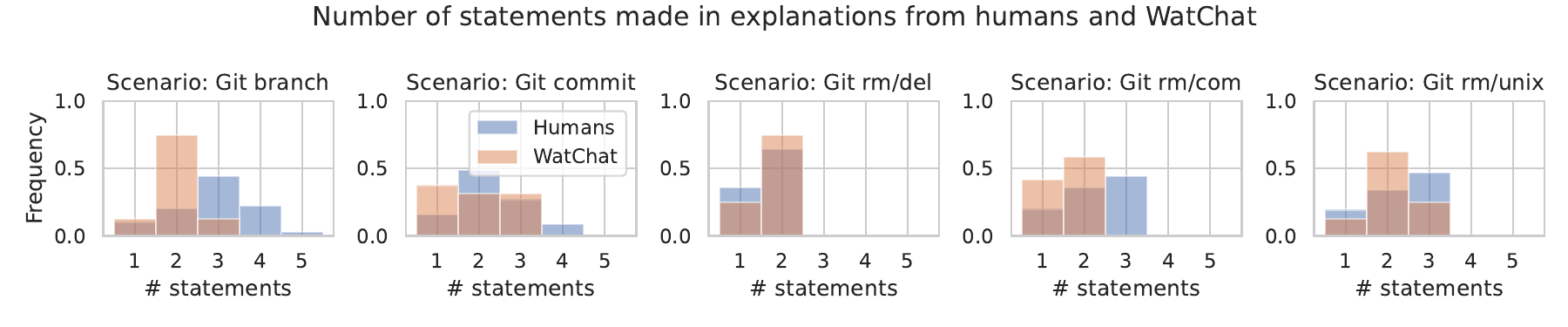}
    \caption{What happens to explanations when the user's mental model is ambiguous (Section~\ref{sec:git-analysis})? In aggregate, human-authored explanations address a variety of statements for each scenario (top); however, each individual human-authored explanation typically contains only a subset of those statements (bottom). This suggests that different people inferred different erroneous mental models and addressed different misconceptions. \watchatgit\ exhibits similar behavior, though in many scenarios its explanations contain fewer statements than human-written explanations. This in turn suggests new misconceptions to add to our system. (See note on error bars in Figure~\ref{fig:summary}.)}
    \label{fig:git-num-stmt}
\end{figure*}

\subsubsection{Length and correctness of explanations} Figure~\ref{fig:summary} shows that human-written explanations, as well as \watchat's, tend to be relatively succinct (humans average $272\pm 186$ characters, while \watchat\ averages $297\pm63$ characters). On the other hand, GPT-4's explanations were much longer (averaging $1343 \pm 379$ characters). By a two-tailed $t$-test, GPT-4's explanations were significantly longer than humans' and \watchat's ($p \ll 0.01$). Humans' and \watchat's explanations were not significantly different in length ($p = 0.31$).

\watchat\ is designed such that explanations are correct by construction, and indeed $0\%$ of explanations were rated as incorrect. Overall, $6\%$ of human-written explanations were rated incorrect, and $35\%$ of GPT-4-written explanations were rated incorrect. By a two-tailed $t$-test, GPT-4's explanations were incorrect significantly more often than humans' ($p \ll 0.01$).


\subsubsection{Meeting Miller's desiderata}\label{sec:js-analysis} Figure~\ref{fig:js-summary} shows how explanations in the JavaScript domain depended on the scenario given. We found that \watchatjs\ generally matched human sensitivity to the scenario, whereas GPT-4 was often ambivalent to them.

For example, consider example pair ``C'' (bottom row), where a slight modification of the array being indexed causes people and \watchat\ to switch from explaining zero-indexing to explaining lexicographic ordering. This shows that people and \watchatjs\ are both social and selective in their explanations (they select what to explain, by reasoning about what the question-asker needs). GPT-4, however, has a strong tendency to explain lexicographic ordering no matter what---and about half the time fails to explain zero-indexing even when it is relevant.

Similarly, consider example pair ``B'' (middle row), where asking about the same program but changing \emph{only} the expectation causes people to reliably switch from explaining that \lstinline{typeof(null)} is \lstinline{"object"} to explaining about equality-by-reference. This shows that people and \watchat\ are contrastive in their explanations (they take into account the question-asker's expectation). GPT-4, on the other hand, has a strong tendency to explain about \lstinline{typeof(null)} no matter what.

Finally, consider example pair ``A'' (top row), where introducing the ``red herring'' of the truthy empty object does not distract \watchatjs\ or most people from the essence of the program's confusing behavior. This again shows that people and \watchat\ are selective in their explanations. On the other hand, GPT-4 consistently needlessly explains about the empty object's truthiness when the ``red herring'' is introduced.

Example pair ``A'' also shows some differences between explanations produced by humans and explanations produced by \watchatjs. For example, as shown in the leftmost plot in the row, \watchatjs\ did not explicitly state that \lstinline{+} coerces non-numerical arguments to string, because from \watchatjs's perspective, a person who expects \lstinline{"[2,undefined,2]"} must already know about coercion to have that expectation. Nonetheless, people do explicitly make this statement relatively often, suggesting that \watchatjs\ could be improved by explicitly stating this fact when discussing coercion caused by \lstinline{+}.

Similarly, as shown in the third plot in the row, \watchatjs\ explicitly states that square brackets are not present when arrays are coerced to string. However, people often do not make this statement. We speculate that this is because people do not notice this feature, especially in light of the more prominent absence of a token like \lstinline$null$ or \lstinline$undefined$. In this case, we believe \watchatjs\ is right to explicitly make this statement.

\subsubsection{Choosing among multiple explanations}\label{sec:git-analysis}
The analysis above, using JavaScript scenarios, focused on cases where the user's expectation gives a strong indication as to their likely misconceptions. Next, we use Git scenarios to consider cases where many different erroneous mental models could lead to the same expectation-violation (namely, an error where none was expected). Because GPT-4's explanations were consistently wrong for most Git scenarios, we excluded GPT-4's explanations from these analyses.

Our key question is whether \watchatgit\ matches human intuitions about selectivity in explanation in the case of multiple possible erroneous mental models leading to the same expectation violation. \watchatgit's behavior in these cases is to consider these possible mental models one at a time, ranked by an estimate of likelihood, and generate explanations independently and selectively with respect to each possible mental model. Is this what people do? That is, do people intuitively select a single likely erroneous mental model to debug, or do they instead ``cover their bases'' and give a merged explanation that non-selectively gives all of the potentially-relevant information at once?

The top row of Figure~\ref{fig:git-num-stmt} shows that for each scenario, in aggregate human responses made a variety of statements. However, as shown in the bottom row, individual explanations often only addressed a subset of statements. For example, in the ``Git branch'' scenario, in aggregate human-written explanations covered five statements (top row, leftmost plot), but each individual explanation only addressed an average of three statements (bottom row, leftmost plot). This suggests that individuals are explaining selectively with respect to the mental model they infer, and that different people infer different mental models. This indeed accords with \watchatgit's approach, which is to infer multiple possible erroneous mental models and issue an explanation for each possibility.

\subsubsection{Differences between humans and \watchatgit}
The bottom row of Figure~\ref{fig:git-num-stmt} also shows that \watchatgit\ often under-explains compared to humans. For example, in the ``Git branch'' scenario, \watchatgit\ only addresses an average of \emph{two} statements across all possible explanations it generates. This suggests that \watchatgit's explanations differed from humans' in some important ways.

For example, a closer look at the ``Git rm/unix'' scenario reveals that people often explain that uncommitted changes are ``carried over'' when switching branches, a statement that \watchatgit\ never made for this scenario (this statement is represented by the third bar, blue, in the rightmost plot in Figure~\ref{fig:git-num-stmt}). After noticing this, we realized that \watchatgit\ was missing a misconception. We had originally included a misconception that \lstinline{git checkout} saves and overwrites uncommitted-but-staged changes when changing branches (instead of carrying them over to the new branch) --- but crucially, we had \emph{not} included the analogous misconception for \emph{unstaged} changes. After making a one-line change to \watchatgit\ to account for this, we found that \watchatgit\ indeed explains about uncommitted changes ``carrying over'' for this scenario.



Similarly, a closer look at the ``Git branch'' scenario shows that people often make the statement ``switching to branch \lstinline{main} when you are already on \lstinline{main} is a no-op,'' which \watchatgit\ never made (this statement is represented by the fourth bar, pink, in the leftmost plot in Figure~\ref{fig:git-num-stmt}). After noticing this, we updated \watchatgit's explanation generation to produce such a statement about \lstinline{git checkout} when appropriate. (We had previously overlooked the fact that a false belief about the current branch is relevant to the behavior of \lstinline{git checkout} in the special case that the source and target branches are the same.)


The changes described in the above paragraphs took us only around 10~minutes each to implement. Thus, even though \watchatgit's inventory of misconceptions is currently far from complete, we are optimistic that its design allows for growth over time.

\section{Conclusion}

In this paper, we described a new framework, \watchat, for explaining unexpected behavior in programs. Our framework builds on empirical cognitive science work on explanations: it infers misconceptions that would cause the user to ask ``why?'' (or exclaim ``WAT!'') in response to the given program's output, and then generates an explanation to debug those misconceptions.
We showed how to instantiate our framework in two domains, JavaScript and Git, and we evaluated the two resulting systems by comparing them to 200 human-written explanations, as well as explanations generated by a state-of-the-art language model.

Taking a step back: in this paper, we imagine a future where our programming environments act not only as tools, but as collaborators. The \watchat\ framework begins to demonstrate how studying the cognitive science behind cooperative social interaction can help us design thoughtful, human-like interlocutors from first principles.

\section*{Acknowledgements}
We thank Sorawee Porncharoenwase for help using Rosette, Yoni Friedman for help coding responses, the many anonymous participants in our study, Shriram Krishnamurthi and Kuang-Chen Lu for a thought-provoking discussion about ``misinterpreters,'' Jonathan Zong, Sam Estep, and Manya Bansal on advice on writing this paper, and the attendees of the PLATEAU~2024 workshop for a series of lovely conversations about these early-stage ideas. This research was funded by NSF grants \#CCF-1231216, \#CCF-1723445 and \#2238839, and ONR grant \#00010803. Additionally, KC was supported by the Hertz Foundation, the Paul and Daisy Soros Fellowship, and an NSF Graduate Research Fellowship under grant \#1745302. KMC was supported by a Marshall Scholarship, King's College Cambridge, and the Cambridge Trust. AW  acknowledges  support  from  a  Turing  AI  Fellowship  under grant  EP/V025279/1,  ELSA,  and  the Leverhulme Trust via CFI.



\bibliography{refs}

\newpage 
\newpage

\addtocontents{toc}{\protect\setcounter{tocdepth}{-1}}
\appendix

\tableofcontents

\addtocontents{toc}{\protect\setcounter{tocdepth}{1}}
\section{\watchat\ for JavaScript}\label{sec:watchatjs}

\begin{table*}[]
    \centering
    \begin{tabular}{p{0.25\linewidth}p{0.35\linewidth}p{0.35\linewidth}}
    \toprule
       \textbf{\watchat\ framework element} & \textbf{\watchatjs\ implementation} & \textbf{\watchatgit\ implementation} \\
    \midrule
       Program $p$ & A JavaScript expression & A sequence of Git and Bash commands \\ \addlinespace
       True semantics $\Sigma$ & Standard ECMAScript semantics, for which we wrote a custom interpreter & The official Git implementation, which we query directly via GitPython \\ \addlinespace
       True result $r$ & The result (a JavaScript value) that is produced by evaluating $p$ & The final repository state and the command-line output produced by running $p$ \\
    \midrule
       User's (erroneous) mental model $\widetilde{\Sigma}$ & \multicolumn{2}{c}{\emph{A ``misinterpreter'' created by introducing bugs into the implementation of $\Sigma$.}} \\ \addlinespace
       User's expected result $\widetilde{r}$ & \multicolumn{2}{c}{\emph{The result produced by running $p$ under the erroneous semantics $\widetilde{\Sigma}$.}} \\ \addlinespace
       Synthesis query & \multicolumn{2}{c}{\emph{Find a $\widetilde{\Sigma}$ under which the user would expect $p$ to produce some $\widetilde{r} \not= r$.}} \\
    \midrule
       Synthesis search procedure & Constraint solver & Parallelized enumerative search \\
    \bottomrule
    \end{tabular}
    \caption{A summary of the key elements of the \watchat\ framework, and a comparison between how those elements are realized in the implementations of \watchatjs\ and \watchatgit. The three rows in the middle layer encompass both systems.}
    \label{tab:comparison}
\end{table*}

\begin{figure*}[t]
\includegraphics[width=\linewidth,page=2,trim={0 18cm 0 0},clip]{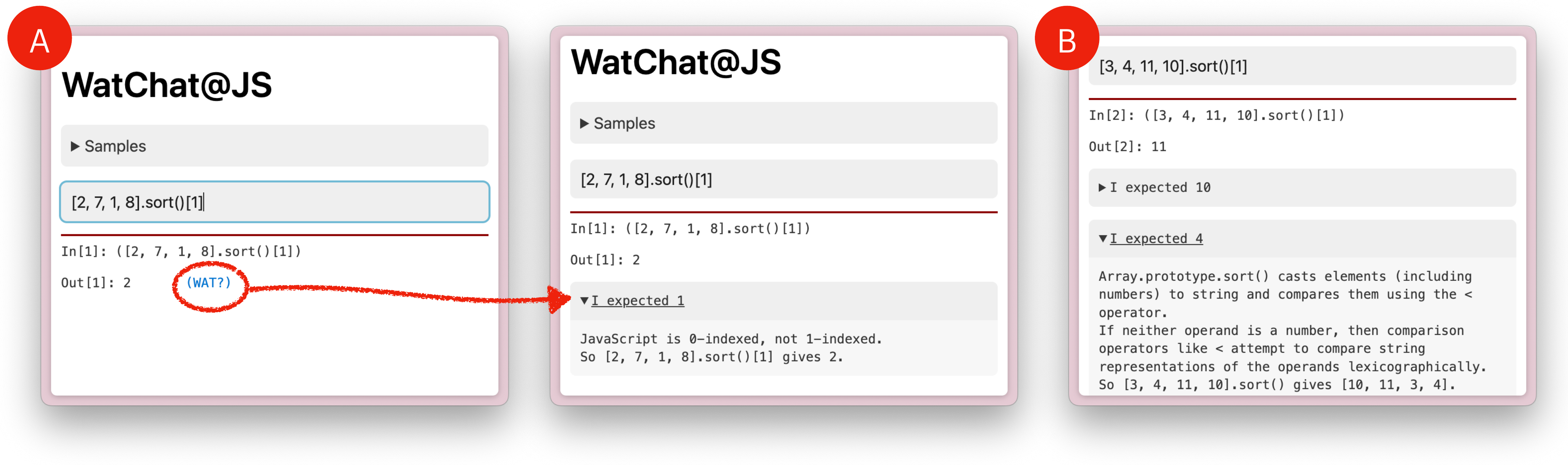}
\caption{\watchatjs's user interface. The user enters an expression in a REPL, which offers a ``WAT?'' button alongside the output. When clicked, \watchatjs\ clarifies what the user expected, and offers an explanation. Panel ``B'' shows a case where multiple possible explanations were found. (The dialogs are closed by default and only open when clicked.)}\label{fig:ui}
\end{figure*}

Our first instantiation of the \watchat\ framework is in the JavaScript domain: we seek to give good explanations in response to ``why?'' questions about type coercion in JavaScript, like the one in the introduction of this paper.

\subsection{Implementation details}

We started by implementing an interpreter for a subset of the ECMAScript 2024 specification \citep{ecmascript2024specification} that covers the ``essence'' of type coercion in JavaScript (inspired by \citet{guha2010essence} and \citet{ park2020jiset}).\footnote{The subset we implemented includes the primitive types \lstinline{Undefined}, \lstinline{Null}, \lstinline{Boolean}, \lstinline{String}, \lstinline{Number}, and \lstinline{Object} (with support for \lstinline{Array} exotic objects), but not \lstinline{Symbol} or \lstinline{BigInt}. We support expressions involving the operators \lstinline{typeof}, \lstinline{!}, ternary \lstinline{?:}, \lstinline{===}, \lstinline{==}, unary \lstinline{+}, binary \lstinline{+}, unary \lstinline{-}, binary \lstinline{-}, \lstinline{<}, \lstinline{>=}, \lstinline{&&}, \lstinline{||}, and \lstinline{??} on these types. Finally, we support indexing arrays, strings, and objects with subscripts \lstinline{[]}, and we support the \lstinline{.sort()} method on Arrays.} As we read the specification, we marked parts that we found unintuitive or surprising. Then, based on our notes, we introduced $N=32$ potential bugs (representing potential misconceptions) in our interpreter. These bugs/misconceptions are listed in Table~\ref{tab:misconceptions}. (Section~\ref{sec:enumeration} further discusses the logistics of enumerating and implementing these bugs.)

We designed our interpreter so that each misconception can be toggled on and off independently with a boolean flag. Each of the $2^N$ possible subsets of bugs $M$ represents one ``misinterpreter,'' i.e.\  one possible erroneous mental model that a user might have about JavaScript's semantics. We define $\Sigma_M$ to be the semantics given by the (mis)interpreter containing the bugs in set $M$. Of course, $\Sigma_{\{\}}$ corresponds to standard ECMAScript semantics because there are no bugs present. We will continue to refer to standard ECMAScript semantics with the unadorned $\Sigma$.

As an example, below is our implementation of the \lstinline{typeof} operator, as described in \S13.5.3 of the ECMAScript specification. This definition includes two potential misconceptions: \lstinline[language=rosette]{typeof-null-is-null}, which says that \lstinline{typeof(null)} is \lstinline{"null"} (it is actually \lstinline{"object"}), and \lstinline[language=rosette]{typeof-array-is-array}, which says that \lstinline{typeof} returns \lstinline{"array"} for Arrays (it actually returns \lstinline{"object"}).
\begin{lstlisting}[language=rosette]
(define (sem-typeof M val)
  (cond
    [(and (mis-typeof-null-is-null? M)
          (js-null? val))
     (js-string "null")]
    [(js-null? val) (js-string "object")]

    [(js-undefined? val) (js-string "undefined")]
    [(js-boolean? val)   (js-string "boolean")]
    [(js-number? val)    (js-string "number")]
    [(js-string? val)    (js-string "string")]

    [(and (mis-typeof-array-is-array? M)
          (js-array? val))
     (js-string "array")]

    [(js-object? val) (js-string "object")]))
\end{lstlisting}
Let us briefly review our notation using this example: if we let $M=\{\texttt{typeof-null-is-null}\}$, and if we consider the program $p=$ \lstinline{typeof(null)+typeof([])}, then a programmer who has the misconceptions in set $M$ would expect program $p$ to produce the output $\llbracket p \rrbracket_{\Sigma_M} = $ \lstinline{"nullobject"}. Of course, under standard ECMAScript semantics that program actually produces the output $\llbracket p \rrbracket_\Sigma=$ \lstinline{"objectobject"}.

Finally, we defined a probability distribution over these misinterpreters. We assume each misconception $m_i$ is independently present with probability $q_i$, so the probability of misinterpreter $M$ is $P(M) = \prod_{m_i \in M} q_i$. For now, we set all $q_i$ equal. This has the effect of placing a higher prior belief on misinterpreters with fewer bugs. Naturally, this is a very coarse estimate of the distribution of erroneous mental models a user might have---in Section~\ref{sec:discussion}, we discuss in depth the limitations of this estimate.

We can now formally state the problem of inferring the user's erroneous mental model. Given program $p$, we seek the set $M$ that maximizes $P(M)$ subject to the constraint $\llbracket p \rrbracket_{\Sigma_M} \not= \llbracket p \rrbracket_{\Sigma}$. To find this set in practice, we use solver-aided program synthesis, facilitated by the Rosette platform \citep{torlak2013growing, torlak2014lightweight}. We use standard symbolic execution techniques \citep{bodik2017domain, chandra2017bonsai} to compile our JavaScript misinterpreters into a set of constraints, and then we use the Z3 solver \citep{de2008z3} to efficiently solve those constraints for $M$. Z3 easily finds solutions to such queries within a few milliseconds.

Now that we have inferred the user's erroneous mental model, the last step is to emit a textual explanation that addresses the user's misconceptions. To do this, we trace through the execution of $p$ under standard semantics ($\Sigma$), and \emph{selectively} only report the execution steps that would be affected by a misconception in $M$. Each time we report an execution step to the user, we also print a message that addresses the respective misconception.

\subsection{Results}\label{sec:js-examples}

Now, we will work through a few example programs whose behavior \watchatjs\ explains in interesting ways. We will start with a case with a single relevant misconception (\S\ref{sec:js-eg-warmup}); then, we will move on to cases where the user might have multiple possible misconceptions leading to different expectations (\S\ref{sec:js-eg-multiple}), cases where the user may have multiple misconceptions at once (\S\ref{sec:js-eg-together}), and finally a case that combines all of these situations (\S\ref{sec:js-eg-final}). We have labeled these programs for easy reference in subsequent sections.

The programs we show here are inspired by and representative of the kinds of programs that users often ask about on platforms like Stack~Overflow: not necessarily real-world code, but rather boiled-down minimal reproducible examples \citep{so2014mwe, lippert2014how} that people typically share in their questions (e.g.\ see \cite{so2010lt, so2011array, so2009nan, so2011zero}).

\subsubsection{Warm-ups}\label{sec:js-eg-warmup}
As a first example, consider the program
\begin{lstlisting}[caption={Program for scenario C/idx},captionpos=b]
console.log( [2, 7, 1, 8].sort()[1] );
\end{lstlisting}
When the user types this into \watchatjs, they see that it returns \lstinline{2}. Additionally, a button labeled ``WAT?'' appears on the side. If the user did not expect \lstinline{2}, they can click the button to reveal an explanation. When the button is clicked, we pose a query to Z3 and receive the solution set $M=\{\texttt{0-indexed}\}$, i.e.\ the singleton set containing only misconception~\#11 in Table~\ref{tab:misconceptions}. Under semantics $\Sigma_M$, the given program produces \lstinline{1} instead of \lstinline{2}. Thus, our system asks, ``Did you expect \lstinline{1}?'' and if the user agrees, then it explains that {\color{gray} ``JavaScript is 0-indexed, not 1-indexed. So, \lstinline{[2, 7, 1, 8].sort()[1]} gives \lstinline{2}.''} This is shown in panel ``A'' of Figure~\ref{fig:ui}.

That was a relatively straightforward explanation. Now consider a variation on the program above, changing the array values and index to be \lstinline$[3, 4, 11, 10].sort()[0]$. Somewhat surprisingly (even to experts), this returns \lstinline{10}. If the user clicks ``WAT?'' \watchatjs\ asks, ``Did you expect \lstinline{3}?'' If the user agrees, then \watchatjs\ explains that {\color{gray} ``"\lstinline$Array.prototype.sort()$ casts elements (including numbers) to string and compares them using the \lstinline$<$ operator.
If neither operand is a number, then comparison operators like \lstinline$<$ attempt to compare string representations of the operands lexicographically. 
So \lstinline$[3, 4, 11, 10].sort()$ gives \lstinline$[10, 11, 3, 4]$.  
So \lstinline$[3, 4, 11, 10].sort()[0]$ gives \lstinline$10$."''}

Notice that these two programs are structurally identical to one another: a complete trace of the interpreter would show the same pattern of behavior on both programs, differing only in the particular numerical values being sorted and indexed. \watchatjs, however, gives very different explanations of the behavior of the two programs. The first explanation addresses 0-indexing, whereas the second explanation addresses the peculiarities of sorting. Unlike the first explanation, the second explanation shows the intermediate sorted list (because the asker likely expects the wrong sorted list). However, it does \emph{not} address 0-indexing (because there is no evidence that the asker has a misconception about indexing). These two examples jointly illustrate how our system is \emph{selective} (designed to only surface relevant information), \emph{social} (designed to reason about the asker's mental state), \emph{contrastive} (designed to reason with respect to the asker's expectation), and \emph{causal} (designed to report causally-relevant features of the semantics, not just a descriptive account of the program trace).

\subsubsection{Multiple possible explanations}\label{sec:js-eg-multiple}
\watchatjs\ is robust to situations where multiple possible explanations exist. To check if there are alternate explanations, we add an additional distinctness constraint to the solver and query again. Given an existing solution $M_1$, we add the constraint that $\boxed{\llbracket p \rrbracket_{\Sigma_{M_1}} \not= \llbracket p \rrbracket_{\Sigma_M} \vee M_1 \not\subseteq M}$. In other words, the new solution must either produce a fresh expected output, or give a nontrivially different new misconception set for the same expected output (i.e.\ not a superset). When there are multiple competing explanations, we can ask the user a clarification question (``Did you expect ...?'') to disambiguate between them.

To illustrate this, consider the same program as above, but now indexed at \lstinline{1} instead of \lstinline{0}:
\begin{lstlisting}[caption={Program for scenario C/lex},captionpos=b]
console.log( [3, 4, 11, 10].sort()[1] );
\end{lstlisting}
This program outputs \lstinline{11}. If a user finds this surprising, then they could have either or both of the two misconceptions referenced above. Thus, for this program, \watchatjs\ asks a clarification question to determine the misconception the user has: ``Did you expect \lstinline{10}, \lstinline{3} or \lstinline{4}?'' Depending on the user's response, \watchatjs\ explains about 0-indexing, lexicographic comparison, or both respectively. This is shown in Panel ``B'' of Figure~\ref{fig:ui}.

Here is a more sophisticated example of this phenomenon, inspired by interesting cases found in the WTFJS~\citep{wtfjs} collection. Consider the following program:
\begin{lstlisting}[caption={Program for scenarios B/null and B/ref},captionpos=b]
function f(x, y, t, u) {
  return (
    t == u ? "hello" : typeof(x)+"/"+typeof(y)
  );
}
console.log( f(null, undefined, {}, {}) );
\end{lstlisting}
If the user runs this program in \watchatjs\ and presses ``WAT,'' \watchatjs\ responds by asking whether they expected \lstinline{"hello"} or \lstinline{"null/object"}.
If the user expected \lstinline{"hello"}, then \watchatjs\ explains that {\color{gray} ``the operators \lstinline{==} and \lstinline{===} compare objects and arrays by reference, not by value. 
So \lstinline${} == {}$ gives \lstinline{false}.  
So your program gives \lstinline{"object/undefined"}.''}

However, if the user expected \lstinline{"null/object"}, then \watchatjs\ instead explains that {\color{gray} ``\lstinline$null$ has type \lstinline$"object"$ (not \lstinline$"null"$, as you might expect). 
So \lstinline$typeof(null)$ gives \lstinline$"object"$.  
So \lstinline$(typeof(null) + "/")$ gives \lstinline$"object/"$.  
So \lstinline$((typeof(null) + "/") + typeof(undefined))$ gives \lstinline$"object/undefined"$.  
So your program gives \lstinline$"object/undefined"$.''}. Notice that \watchatjs\ makes no reference to equality-by-reference in this case, and that \watchatjs\ only shows the \emph{relevant} portion of the program trace (in particular, eliding that \lstinline${}=={}$ gives \lstinline$false$).

\subsubsection{Multiple misconceptions at once}\label{sec:js-eg-together}
Next, consider the following program:
\begin{lstlisting}[caption={Program for scenario A/str},captionpos=b]
console.log("Answers:" + [true, null] + [false]);
\end{lstlisting}
This program outputs \lstinline$"Answers:true,false"$ --- notice that \lstinline{null} seems to ``disappear.'' When the user clicks the ``WAT'' button, \watchatjs\ surfaces multiple possible expectations. The most interesting expectation is \lstinline$"Answer:[true,null][false]"$. In this case, \watchatjs\ explains, {\color{gray} \lstinline{null} is printed as empty string when arrays are cast to string.
When converted to string, arrays don't have square brackets (\lstinline{[]}) around them. 
So \lstinline$[true, null]$ gives \lstinline$"true,"$.  
So \lstinline$("Answers:" + [true, null])$ gives \lstinline$"Answers:true,"$.  
So your program gives \lstinline$"Answers:true,false"$.
} Notice that in this case, \watchatjs\ detected \emph{two} misconceptions to explain to the user (\lstinline{null} and square brackets).
The other two expectations surfaced by \watchatjs\ are cases where the user only has one of these two misconceptions. If the user expected \lstinline$"Answers:true,nullfalse"$ or \lstinline$"Answers:[true,][false]"$, then \watchatjs\ only mentions \lstinline{null} or the square brackets, respectively.

\subsubsection{All together now}\label{sec:js-eg-final}
As a last example that combines everything we have seen so far, consider what happens if we create a program that is similar in spirit to the one above, but slightly more complex.
\begin{lstlisting}[caption={Program for scenario A/truthy},captionpos=b]
function f(x, y) { return x && ("" + y); }
console.log( f({}, [2,undefined,2]) );
\end{lstlisting}
This program outputs \lstinline$"2,,2"$. To explain this behavior to a surprised user, \watchatjs\ asks them to select from an array of possible expectations, which includes \lstinline{false} along with possible alternate string values like \lstinline{"[2,undefined,2]"}, \lstinline$"2,undefined,2"$, and \lstinline$"[2,,2]"$. In the first case, \watchatjs\ explains that \lstinline${}$ is ``truthy'' in JavaScript. In the second and more interesting case, \watchatjs\ explains that {\color{gray} \lstinline{undefined} is printed as empty string when arrays are cast to string.
When converted to string, arrays don't have square brackets (\lstinline{[]}s) around them. 
So \lstinline{[2, undefined, 2]} gives \lstinline{"2,,2"}.} Explanations for the subsequent cases omit information about the square brackets and \lstinline{undefined}, respectively. Importantly, notice that none of these latter explanations make any mention of ``truthiness'' of the empty object, because the user's expectation implies that they already understand why a string is printed.
\addtocontents{toc}{\protect\setcounter{tocdepth}{1}}
\section{\watchat\ for Git}\label{sec:watchatgit}

\begin{figure}
    \centering
    \includegraphics[width=\linewidth,,trim={0.5cm 0 0.5cm 0},clip]{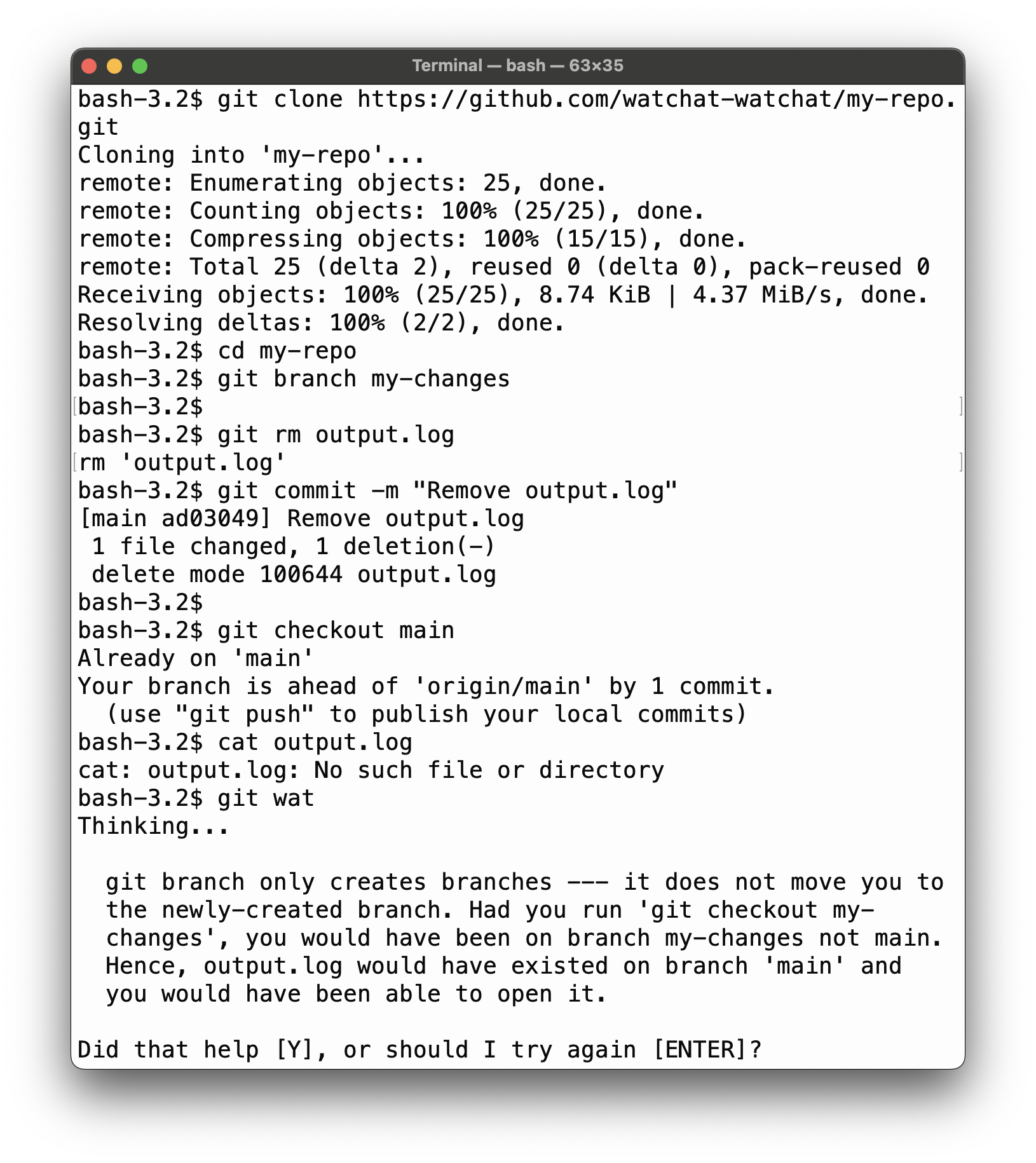}
    \caption{\watchatgit's user interface. The user executes Git and Unix commands in their shell as usual. If they are surprised at any point, they run the \lstinline{git wat} command, which calls \watchatgit. \watchatgit\ reads the user's shell history, analyzes it, and produces a ranked list of explanations, which are shown one at a time.}\label{fig:git-interface}
\end{figure}

Now, we will apply the \watchat\ framework to a domain very different from JavaScript: explaining unexpected behavior of the Git version control system.

Despite being widely-used across the software industry, Git has a reputation for being extremely difficult to learn and work with. For example, in a well-known webcomic, \citet{munroe2015git} lampoons Git by depicting how lay users typically memorize commands and rely on a surface-level understanding to accomplish day-to-day tasks. Studies consistently show that users typically dread using Git, worried that they will make a mistake and irrecoverably lose their work \citep{yang2022developers, church2014case}.

What makes Git so daunting? Many have argued that the difficulty in understanding Git lies in acquiring the right mental model. For example, \citet{perez2013s} critique Git's usability on the basis of its unnecessarily complicated conceptual model, using it to propose a simpler redesign \citep{de2016purposes}. Similarly, \citet{milliken2021behavioral} and \citet{isomottonen2014challenges} identify ``lacking a mental model'' as a major barrier to Git's usability. Perhaps most vividly, in a recent blog post \citet{evans2023git} discusses why people have trouble using branches in Git, outlining ``an intuitive mental model I think many people have
how git actually represents branches internally, how the `intuitive model' and the real way it works are actually pretty closely related,
some limits of the intuitive model and why it might cause problems.''

With all of this in mind, we decided to apply the \watchat\ framework to create a tool that can answer ``why?'' questions about unexpected behavior in Git. Here, we consider the ``program'' $p$ to be a sequence of Git and Unix commands executed by the user on the command line. Our user interface to \watchatgit\ is straightforward: it is a command-line program, \lstinline{git wat}, that users can execute at any point (see Figure~\ref{fig:git-interface}). \watchatgit\ reads and parses the user's recent command-line history to obtain $p$, generates possible explanations, and then prints them directly to the terminal.

\subsection{Implementation details}

At a high level, our approach to implementing \watchatgit\ is no different from our approach to implementing \watchatjs. First, we carefully read a classic Git textbook \citep{chacon2014pro} and compiled a list of 25 potential misconceptions as we came across them (see Table~\ref{tab:git_misconceptions}). Next, we implemented ``misinterpreters'' that counterfactually simulate what a user's commands would do based on their (erroneous) mental model. Finally, we implemented an algorithm that uses our misinterpreters to explain an expectation violation (``WAT!'') by searching over misconceptions to infer the user's erroneous mental model.

We implemented our misinterpreters in Python using the GitPython API \citep{gitpython}. We overrode a key subset of Git's subcommands (namely \lstinline$add$, \lstinline$commit$, \lstinline$rm$, \lstinline$mv$, \lstinline$checkout$, \lstinline$branch$, \lstinline$clone$, \lstinline$reset$, \lstinline$revert$, \lstinline$diff$, and \lstinline$merge$), as well as several common Unix commands (\lstinline$cat$, \lstinline$echo$, \lstinline$cd$, and \lstinline$rm$) in order to change their behavior based on the current set of misconceptions the user has. This allows us to ``misinterpret'' a user's command-line history. For example, our ``misinterpreter'' for the \lstinline$git branch$ command takes in not only the name of the target branch, but also the set of misconceptions the user has. It begins as follows:
\begin{lstlisting}[language=Python]
def git_branch(repo, name, misconceptions): 
    repo.git.branch(name)
    if 'branch_moves' in misconceptions:
        repo.git.checkout(name)
\end{lstlisting}
Here, \lstinline$'branch_moves'$ is the name of the misconception that the command \lstinline$git branch name$ not only creates a fresh branch called \lstinline$name$, but also moves to that branch. If that misconception is activated, then our patched version of \lstinline$git branch$ will have that erroneous behavior.

Now, given a repository, a sequence of commands $p$, and a counterfactual semantics $\widetilde{\Sigma}$ (expressed as a set of active misconceptions), we can evaluate $\llbracket p \rrbracket_{\widetilde{\Sigma}}$, i.e.\ simulate what command-line output a user would expect, by (1)~creating a temporary copy of the repository, (2)~executing the given commands with the given misconceptions, and (3)~capturing the final state of the temporary repository, as well as whatever information is printed to standard output.

The next step is to infer the user's misconceptions given an expectation violation. To automate this process, we need a way to detect expectation violations automatically. In \watchatjs, this was as easy as comparing the program's true output to the user's expected output under their erroneous mental model. With \watchatgit, however, the situation is a little more complicated. In principle, the output of a sequence of commands can be fully characterized by the state of the repository and filesystem after the commands are run, as well as any intermediate information printed to the user's terminal. However, very little of this state is likely to be relevant to the expectation violation. For example, the repository and filesystem state are only observable insofar as the user runs commands to inspect them. Furthermore, because Git often prints quite complex messages to the terminal, users may not even attend to everything that is displayed: for example, they may miss the name of the current branch in the output of \lstinline{git status}, or they might simply ignore or forget information they do not understand.

Thus, for the purposes of \watchatgit\ we explicitly support only the following misconceptions: (a)~a Git command produced an error, even though no error was expected, (b)~a file was displayed (e.g.\ via \lstinline{cat} or \lstinline{vim}) and contained unexpected contents, (c)~an attempt to access a file failed because the file unexpectedly did not exist, (d)~the command \texttt{git diff} produced unexpected output.
Of course, this definition of expectation violation is imperfect: it might raise false positives (e.g.\ a user prints a file's contents but does not notice that it is different than expected), as well as false negatives (e.g.\ a user intentionally raises an error, perhaps to confirm that a file has been deleted). In the future, we plan to allow for a richer space of possible expectation violations, perhaps by soliciting user input in natural language. However, for now we consider this set as representative of the most alarming ``WAT!'' scenarios users face.

When the user observes an expectation violation of this sort, we infer their erroneous mental model by computing a subset of misconceptions that would explain the expectation violation. Recall that the implementation of \watchatjs\ used a constraint solver to perform this search efficiently. That was possible because we created an explicit encoding of JavaScript semantics into a set of SMT constraints, via symbolic execution of our custom JavaScript interpreter. In contrast, \watchatgit\ uses an existing ``off-the-shelf'' Git client, which is not directly compatible with symbolic execution and constraint solving. Hence, \watchatgit\ instead uses a simple enumerative search: we enumerate sets of misconceptions in increasing order of size, and test each candidate mental model to see if it would account for the user's expectation violation. Because Git operations are typically bottlenecked by filesystem operations, this procedure benefits greatly from parallelization, and can therefore run in a reasonable amount of time for an interactive tool. The examples below all run within 20 seconds on a 16-core MacBook Pro.


As with \watchatjs, if more than one erroneous mental model can account for the user's expectation violation, \watchatgit\ ranks them by posterior probability. \watchatgit\ starts with a prior analogous to the one we defined in \watchatjs: we assume that each misconception is present or absent independently with some probability. In \watchatgit, we additionally coarsely estimate the \emph{likelihood} of the user writing program $p$ conditioned on a given mental model $\widetilde{\Sigma}$. This is motivated by the fact that in this domain, the program $p$ can itself reveal a lot about the user's mental model. For example, we posit that it is generally likely that the \lstinline{git branch} command (to create a new branch) is followed by \lstinline{git checkout} (to switch to that branch) --- unless, of course, the user has the misconception that \lstinline{git branch} automatically switches you to that branch, in which case it is \emph{un}likely that the user would redundantly (from their perspective) run \lstinline{git checkout} immediately after \lstinline{git branch}. Hence, when inferring $\widetilde{\Sigma}$ for a program $p$ where \lstinline{git branch} is \emph{not} immediately followed by \lstinline{git checkout}, \watchatgit\ gives additional weight to mental models that include the misconception about \lstinline{git branch}.

Now that we have inferred the user's erroneous mental model, the last step is to emit a textual explanation that addresses the user's misconceptions. Just like \watchatjs, \watchatgit\ generates an explanation by tracing the execution of the user's commands under standard semantics ($\Sigma$). Whenever there is a difference between the observed state and the counterfactual state that would have arisen under the user's erroneous mental model, we report a textual description of the difference, along with a message addressing the relevant misconception.

\subsection{Results}\label{sec:git-examples}

As in the previous section, we will now work through several interesting case studies to showcase the kinds of explanations \watchatgit\ gives. As before, we will label each example scenario for easy reference in future sections.

\paragraph{Scenario ``Git branch''}
Consider a user who plans to create a new branch where they intend to delete a spurious output file (e.g.\ a log) from the repository. However, after deleting this file, the user realizes that they need to refer back to it. Here is a sequence of actions such a user might take. First, they clone the repository and enter the freshly-created directory.
\begin{lstlisting}[language=bash]
git clone https://github.com/.../my-repo
cd my-repo
\end{lstlisting}
Next, they create a new branch for their changes.
\begin{lstlisting}[language=bash]
git branch my-changes
\end{lstlisting}
After this setup, they update the contributing guidelines and then delete the log file.
\begin{lstlisting}[language=bash]
git rm output.log
git commit -m 'remove log'
\end{lstlisting}
But then, they realize that they need to refer back to the old log file, so they attempt to reference the old version that should still be on branch \lstinline$main$.
\begin{lstlisting}[language=bash]
git checkout main
cat output.log
\end{lstlisting}
However, at this point they get a scary error: the file \lstinline{output.log} does not exist. Why did they get that error?

A good explanation here is that even though the user \emph{created} the new branch with the \lstinline{git branch} command, they never switched to that branch (even though they might have thought they did). \watchatgit\ thus produces the following explanation: {\color{gray} git branch only \emph{creates} branches --- it does not move you to the newly-created branch. Had you run \lstinline{git checkout my-changes}, you would have been on branch \lstinline{my-changes} not \lstinline{main}. Hence, output.log would have existed on branch \lstinline{main} and you would have been able to open it.}

So far, so good. But what if the user indeed intended to create the new branch \emph{without} switching to it---or perhaps changed their mind and decided to delete \lstinline{output.log} directly on \lstinline{main}? This is a less likely situation, but one that \watchatgit\ can nonetheless anticipate and reason about. In this case, \watchatgit\ infers that a completely different misconception could be at play: that the \lstinline{git rm} command removes the file ``from the repo, not the disk.'' The second-ranked explanation is therefore: {\color{gray} \lstinline{git rm} not only stages the file's deletion, but also deletes the file from your disk. Use the \lstinline{--cached} flag to avoid this. Had you instead run \lstinline{git rm --cached output.log}, you would have had \lstinline{output.log} in your working tree. Hence, \lstinline{output.log} would have existed and you would have been able to open it.} Notice how unlike the previous explanation, this one does not address branching at all. This shows how \watchatgit\ is selective and social in its explanations: only addressing points that are relevant to the user.

In the rest of this section, we will describe how \watchatgit\ handles a variety of variations on this general narrative: deleting \lstinline{output.log} on a branch, and being surprised when \lstinline{cat} fails.

\paragraph{Scenario ``Git rm/del''} As a warm-up, consider what happens if the user actually \emph{does} switch branches to \lstinline{my-changes}, but runs \lstinline{cat} immediately after \lstinline{git rm} (i.e.\ while still on branch \texttt{my-changes}, rather than after committing and switching branches). In this case, \watchatgit\ gives as its top-ranked (and only) explanation the second explanation above, which explains about the difference between \lstinline{git rm} vs. \lstinline{git rm --cached}. Notice again that the explanation does not address branching at all.

\paragraph{Scenario ``Git rm/unix''} Next, consider what happens if while on branch \lstinline{my-changes} the user happens to use the shell built-in command \lstinline{rm} instead of the Git command \lstinline{git rm}. Then, as before, they commit and switch back to \lstinline{main}. In this case, \watchatgit\ explains that the issue was related to the use of the shell's \lstinline{rm} command: {\color{gray} The shell command rm does not tell Git to stage the file's deletion. To have the deletion tracked by Git, you need to use the Git command git rm instead. Had you instead run \lstinline{git rm output.log}, you would have staged changes to output.log. Hence, output.log would have existed on branch \lstinline{main} and you would have been able to open it.} A lower-ranked explanation assumes that the user \emph{does} know the difference between \lstinline{rm} and \lstinline{git rm}, and instead simply forgot to stage the changes.

\paragraph{Scenario ``Git rm/com''} Now suppose that the user did indeed run \lstinline{git rm} as before, but did not run \lstinline{git commit} afterwards. In this case, \watchatgit\ explains that {\color{gray} \lstinline{git rm} only stages the file's deletion; it does not commit the change. Had you instead run \lstinline{git commit} after \lstinline{git rm output.log}, you would have taken a snapshot of the changes to output.log in your staging area. Hence, output.log would have existed on branch \lstinline{main} and you would have been able to open it.}

\paragraph{Scenario ``Git commit''}
Finally, suppose that while on the branch \lstinline{my-changes}, the user also adds a note to the project's contributing guidelines.
\begin{lstlisting}[language=bash]
git checkout my-changes
echo "..." >> CONTRIBUTING.txt
git add CONTRIBUTING.txt
git rm output.log
git checkout main
\end{lstlisting}
Based on this additional evidence, \watchatgit's behavior changes in a subtle way. Based on the user's actions here, it seems possible that they have a higher-level misconception: they might not realize at all that there is a distinction between \emph{staging} changes and \emph{committing} them in Git---perhaps they think \lstinline{add} and \lstinline{rm} directly save snapshots of changes to the current branch. The highest-ranked explanation is thus: {\color{gray} Git has a notion of a ``staging area'' where you add changes to be recorded as part of the next commit. Had you run \lstinline{git commit} after \lstinline{git rm output.log}, you would have taken a snapshot of the changes to CONTRIBUTING.txt and output.log in your staging area. Hence, output.log would have existed on branch \lstinline{main} and you would have been able to open it.}
The second-highest-ranked explanation assumes that the user \emph{does} know the distinction between staging and committing, but does not realize that \lstinline{git rm} and \lstinline{git add} only stage the changes; this explanation is similar to the one for Scenario ``Git rm/com.''

\section{Additional details on evaluation}\label{sec:eval-details}

We allowed participants to use any resources at their disposal, including but not limited to consulting on-line documentation (we provided links to official references) and executing code on their own machines. In particular, for the Git scenarios, we made the example repository available on GitHub so that they could easily reproduce the behaviors shown in the emails. Participants were specifically instructed \emph{not} to use language-model-based AI tools like ChatGPT or Copilot.

Each participant was randomly shown one scenario from each of our three pairs of JavaScript scenarios (in randomly-shuffled order), and then all five Git scenarios (again, in randomly-shuffled order). Hence, we collected $3+5=8$ responses from each participant, for a total of 200 responses.

\subsection{Additional details on human explanation writing task}\label{sec:additional-human-exp}

\subsection{Additional details on coding}\label{sec:coding}

To code the statements made by a given explanation, we first created an inventory of possible statements for each task, and then for each response, we coded which subset of statements were made in the response. To create the inventory of possible statements, one author and one coder not associated with this research independently read all of the responses and independently drafted representative inventories of possible statements for each scenario. Then, they compared their inventories and reconciled differences. The differences were minimal, which speaks to the remarkable consistency of responses we received.

Coders were presented with the explanations grouped by scenario and shuffled within scenario (to be clear, the explanations included not only human responses, but also outputs of \watchat\ and GPT-4, and coders were blind to the source of each explanation). Coders were instructed \emph{not} to code statements prefaced with phrases like ``By the way'' or ``just FYI,'' which explicitly mark information as extra and not part of the core explanation. For each scenario, coders were given the option to indicate that the response contains a crucial statement that was not among the statements listed in the inventory for that scenario (this option was never used). 
Pairwise inter-rater reliability, given by Krippendorff's $\alpha$ \citep{krippendorff2004content, castro2017krippendorff}, was $0.818$ with a bootstrap 95\% confidence interval of $(0.797, 0.840)$, which indicates very good agreement among raters.

\section{Limitations, discussion, and future work}\label{sec:discussion}

\subsection{Threats to validity}\label{sec:eval-ttv}

Our results provide preliminary evidence in support of \watchat's design decisions, but that support is limited by a few threats to validity.

\paragraph{Internal validity}

One confounder in the experiment design is sequencing effects in the human subject study. Participants may change their explanation strategy on later tasks (e.g., getting better with practice, or getting worse with tiredness). The Git scenarios are similar to one another, so a participant's understanding of an earlier scenario may influence their explanations for a later scenario. A few responses even explicitly referenced prior tasks. 

We reduce the influence of sequencing effects in two ways. First, the task order is randomized, so a specific task should not consistently benefit from being earlier or later in the sequence. Second, we emphasize in the instructions that each task is meant to be a query from a different colleague, and so participants should treat each response as an explanation to a new person.

Another threat to internal validity is the online, asynchronous nature of the experiment. Research staff were not present during completion of the experiment, so participants could have engaged in bad faith. However, our qualitative review of all the responses did not find any evidence of such behavior.

\paragraph{External validity}

One threat to external validity is whether the tasks in this study reflect the distribution of tasks in the real world. We hand-selected tasks based on our impression of common ``why?'' questions (Section~\ref{sec:js-examples}), but it is possible that these are not the same scenarios encountered by most users. We combat this threat in part by mining misconceptions from popular learning resources about JavaScript and Git. But this threat is nonetheless an important question for future work, discussed in Section~\ref{sec:discussion}.

Another threat to external validity is whether our elicited GPT-4 explanations are sufficiently representative of LLM capabilities. At the time of writing, GPT-4 is largely considered state-of-the-art~\citep{helm, chatBotArena}, and we expect its explanations to be the best among today's LLMs. Notably, however, we did not conduct any ``prompt engineering'' to elicit different kinds of outputs from GPT-4; humans and the LLM both received the same prompt. A different prompt, parameter setting (e.g., temperature), or a different fine-tuning of the base model could easily elicit at least shorter responses. The response length distribution in Figure~\ref{fig:summary} represents LLM explanations given \emph{human-style prompting} with \emph{default settings} (i.e., how a standard end-user would use a tool like ChatGPT). It is possible that prompt engineering and tweaking settings could potentially produce fewer wrong explanations and/or explanations with statement profiles that better match humans. However, ``prompt engineering'' places substantial burden on the user: those without experience engaging with chatbots may not know the best prompting strategies to employ~\citep{collins2023evaluating}.

\paragraph{Construct validity}

This experiment measures the extent to which machine-generated explanations (\watchat{} or LLM) are consistent with the intuitions of human explainers. This is not the same as measuring the efficacy of the explanations. For example, an alternative experiment design would recruit participants who have misconceptions, provide them explanations, and observe whether the explanation helps the participant: both in completing the task at hand, and in transferring their new knowledge onto new tasks.

\subsection{On the explicit enumeration of misconceptions}\label{sec:enumeration}
There are many forms of unexpected behavior our systems currently cannot handle. We of course cannot explain unexpected behavior that falls outside the subsets of JavaScript and Git for which we have explicitly collected misconceptions. For example, \watchatjs\ cannot yet explain the subtleties of JavaScript's prototype-based object system, and \watchatgit\ cannot yet explain surprises encountered while running the command \lstinline{git rebase}. More generally, a key limitation of our system is the need to explicitly enumerate and formalize a set of misconceptions, which can have a high up-front implementation cost and lead to ``blind spots'' in explanation.

We are nonetheless optimistic that an explicit enumeration of misconceptions can be relatively easy to read, write, and maintain. Importantly, misconceptions do not need to all be introduced up-front at the same time. Rather---as we observed as we built our systems---misconceptions can easily be added and tweaked as they are discovered (for example, in Section~\ref{sec:git-analysis} we mentioned how it took us just 10 minutes to add a new misconception to \watchatgit). The code snippets in this paper show that misconceptions can typically be implemented in just one to two lines of code. In many ways, then, assembling a catalogue of misconceptions is no different from assembling a catalogue of linter rules, carefully-written warnings (for example, those found in the Elm compiler \citep{czaplicki2015compiler, peijnenburg2016type}), or a list of ``pitfalls''/``sharp edges'' included in the documentation.

In any case, we believe that the drawback of needing to enumerate and implement misconceptions is outweighed by many benefits of an explicit formalization of misconceptions. This is best illustrated by our experiments with using a large language model to elicit explanations: because language models only have implicit ``knowledge'' of language semantics and common misconceptions, they routinely produced incorrect explanations, and behaved in unexpected ways. It is difficult to say which misconceptions a large language model is capable of addressing, and if a new kind of misconception arises, it is difficult to update a large language model to take it into account. Finally, while language models perform surprisingly well on common languages like JavaScript, it is not clear if they would generalize to ``low-resource'' programming languages or APIs without substantial fine-tuning or prompt engineering. In contrast, an explicit enumeration of misconceptions allows us to be sure that we are giving correct, sensible explanations, and allows implementers to flexibly add and modify misconceptions to adapt to changing circumstances.

Another major benefit of an explicit representation of the user's mental model is that it creates the possibility of new applications beyond explanation. In Appendix~\ref{sec:diagnostic}, we show one such early-stage application: automatically verifying and synthesizing diagnostic questions. We were motivated by the fact that educators often use diagnostic questions on assessments to pinpoint students' misconceptions. Such questions are carefully designed so that an incorrect response from a student reveals that they have a particular misconception; collections of such questions are called concept inventories \citep{hestenes1992force}. Unfortunately, brainstorming new diagnostic questions for a concept inventory (e.g.\  when writing a new exam each semester) can be a challenging, time-consuming process \citep{lindell2007they}. Furthermore, there is always the risk that a diagnostic question is imprecise because multiple misconceptions could lead to the same mistake. Thus, high-quality concept inventories are often published as research contributions unto themselves \citep{crichton2023grounded}. It turns out, however, that \watchatjs's infrastructure can be easily applied to \emph{synthesize} diagnostic programs $p$ that precisely isolate a target misconception (Appendix~\ref{sec:diagnostic}).

\subsection{On the estimate of prior distribution}
As we discussed in Section~\ref{sec:git-analysis}, our current systems are limited by their relatively coarse estimate of $P(M)$, the distribution over users' erroneous mental models. In the future, we would like to develop better methods for capturing the distribution of misconceptions. Instead of brainstorming common misconceptions and estimating priors, which is subject to the ``expert blind spot'' \citep{nathan2001expert}, we plan to systematically collect misconceptions based on behavioral studies on real programmers, as previous work has done in other domains \citep{crichton2023grounded, smol2023}. Additionally, we plan to measure the frequencies of these misconceptions empirically, and to model their covariance structure instead of treating misconceptions as independent. Finally, we plan to use additional available sources of information to inform the prior over the current user's mental state. For example, we could take into account previous interactions with the system (either actual or inferred \citep{chandra2024inferring}), whether the user is a beginner or an expert \citep{barik2017developers}, and what other programming languages the user might be familiar with \citep{shrestha2018s}.

\subsection{Beyond semantics}
Our method is currently only applicable to behavior that is unexpected because of misconceptions about language/API semantics. But not all unexpected behavior is due to such misconceptions. For example, we cannot yet explain unexpected behavior caused by misconceptions about syntax, such as operator precedence and automatic semicolon insertion in JavaScript, or shell quoting in Git.
The user might also have misconceptions that are neither about syntax nor semantics. For example, a good explanation for why \lstinline{document.getElementByID("xyz")} fails with an error (``TypeError: undefined is not a function'') is that the user has a misconception about the method name, which is actually called \lstinline{document.getElementById} (lowercase ``d''). But a good explanation for why \lstinline{document.getElemntById("xyz")} fails is that the user made a typing error and misspelled the word ``Element.'' This explanation is not based on a misconception about the language, but rather a false belief about the actual program being executed. Thus, even though these programs nominally both fail with the same error message, their failures are best explained at very different levels. To handle such cases in \watchat, we would ultimately like to jointly infer not only the user's misconceptions and expected output, but also the \emph{intended program} they meant to write. This requires a well-tuned prior distribution over programs that are natural to write. We believe large language models maybe be well-suited for this task. More work is needed to systematically evaluate large language models on explanation tasks, and to design neurosymbolic systems that combine the flexibility of large language models with the guarantees of formal methods.

Finally, we are interested in extending applying \watchat\ framework to explain other forms of surprising program behavior. We are particularly interested in explaining performance regressions by debugging users' mental models about their hardware's performance characteristics. For example, if a user swaps the order of two nested loops in a matrix multiplication routine and observes a surprising slowdown, we would like to automatically explain to them what happened (e.g.\  by explaining that cache hits are important for performance, and that the new program has worse locality). Other interesting domains include page layout (e.g.\ surprising results when positioning figures in CSS or \LaTeX) and security configurations (e.g.\ debugging Cross-Origin Resource Sharing or ``CORS'').

\verb|\appendix| 
\section{Background}\label{sec:background}

Philosophers and cognitive scientists have long debated what makes a good explanation, and a full account is beyond the scope of this paper. We instead direct readers to \citet{miller2019explanation}, who surveys the field for an audience of explainable AI researchers. \citeauthor{miller2019explanation} distills from the literature four key desiderata that characterize human intuition about good explanations:
\begin{enumerate}
\item Good explanations are \textbf{contrastive}, i.e.\ they explain why $Y$ (the ``fact'') is true \emph{instead of} the expected $X$ (the ``foil''). The ``foil'' may or may not be explicit in the ``why?'' question asked \citep{lipton1990contrastive, riveiro2021thats}.
\item Good explanations are \textbf{selective}, i.e.\  they do not exhaustively recount the full causal chain behind an event, but rather a sparse subset chosen pragmatically by the explainer \citep{gerstenberg2020expectations, poesia2022left, lai2023selective}. More broadly, good explanations should obey general principles of efficient communication between intelligent agents  \citep{grice1975logic}.
\item Good explanations are \textbf{causal}, not statistical, in nature. In particular, it is not enough to say ``because Y is more common than X'' \citep{josephson1996abductive} and it is not enough to give the likeliest causal factor \citep{hilton1996mental, mcclure2002goal}.
\item Good explanations are \textbf{social} and interactive, occurring as part of a conversation between two agents in a given context \citep{hilton1990conversational}. Indeed, people often make inferences \emph{about} the social context based on the explanations they are given \citep{kirfel2022inference}.
\end{enumerate}
What would it take to build tools that automatically give explanations that meet these desiderata---that is, tools that match our intuitions about what good explanations look like?
Prior work in PL, HCI, and CS education has led to explanation systems that meet various subsets of these desiderata. 
For example, tools like Tutorons \cite{head2015tutorons} generate standalone explanations of code, which are causal but not contrastive against a reader's expectations.
To take another example, the Whyline \citep{ko2004designing, ko2009finding, ko2010extracting, ko2008debugging} (which inspired \watchat's name) was a seminal system that gave causal, contrastive explanations in the form of traces showing all causally-relevant events that answer ``why?'' or ``why not?'' questions about Java programs. Similarly, Amalgam \citep{nelson2017power} shows logical derivations to answer ``why'' and ``why not?'' questions about models found by Alloy. However, these two methods are not selective (they give the full causal graph) and they are not social (they do not interactively reason over models of their interlocutors). 
Other methods work by flagging uncommon code patterns or common bugs \citep{engler2001bugs, singh2013automated}: that is, they give a statistical account (``this is a common bug'') rather than a causal account of the program's unexpected result (``this bug leads to this behavior because...''). Thus, as \citeauthor{singh2013automated} note, they cannot properly explain bugs caused by conceptual errors.

In this paper we apply recent ideas from cognitive science to build a system that satisfies all of Miller's desiderata at once. Our key idea is that instead of debugging a given \emph{program}, we should infer and debug the user's \emph{mental model} \citep{craik1967nature, johnson1980mental} of the language or API they are using.
We build on a recent line of research that seeks to characterize the nature of cooperative explanation in precise computational terms \citep{chandra2024explanation}. Under \citeauthor{chandra2024explanation}'s view, people (even infants \citep{perez2022violations}) typically ask ``why?'' questions when they are surprised by an outcome. A surprise, unexpected result, or suspicious coincidence \citep{griffiths2007mere} is a cue that our mental model of the world is incorrect or incomplete; that is, that there is an opportunity to learn. The explainer's role is therefore to efficiently infer and debug the asker's mental model. In other words, the explainer is tasked with (1)~estimating what misconceptions the asker has that would cause them to ask ``why?'' and then (2)~fixing the asker's relevant misconceptions, while avoiding redundantly saying anything the asker already likely knows.
\citeauthor{chandra2024explanation} express this view of explanation in a computational model, and, through a behavioral study, they show that their model of explanation predicts human intuitions well across a variety of scenarios in a small ``toy'' experimental domain (planning routes in a grid-world). Now, in this paper, we will show how \citeauthor{chandra2024explanation}'s work can be applied to build real-world tools for programmers.

As we discussed, and as we will formalize in Section~\ref{sec:watchat}, a key facet of our framework is representing the user's erroneous mental model with counterfactual semantics for the language or API they are using.
The abstract idea of representing mental models with debuggable programs (and misconceptions as bugs) is not new. For example, in his book \textit{Mind Bugs}, \citet{vanlehn1990mind} applies this analogy to understand young math students' procedural misconceptions about subtraction, an idea \citet{feldman2018automatic} apply to automatically diagnose K-8 students' misconceptions. More recently, \citet{smol2023} apply a similar idea to formalize student misconceptions about scope and mutation as ``bugs'' in interpreters for a Scheme-like language. Our work goes one step further by automatically inferring and correcting (i.e.~synthesizing and debugging) these mental models. More broadly, our work is related to efforts to apply cognitive theories of mental state inference to design algorithms for automatic tutoring \citep{rafferty2015inferring, rafferty2016faster, anderson1995tutor}, explainable robotics \citep{chakraborti2017plan}, and explainable recommendation systems \citep{tsai2019designing, millecamp2019explain, szymanski2024designing, knijnenburg2012explaining}.


\section{Synthesizing diagnostic programs}\label{sec:diagnostic}
Because \watchatjs\ maintains an explicit representation of the user's mental model, it can be applied to tasks beyond explanation. Here, we give one early-stage motivating example: automatically synthesizing ``diagnostic programs'' or ``concept inventories.''

The infrastructure behind \watchatjs\ can be used to synthesize fresh diagnostic programs that are optimized to pinpoint a given misconception. Given a specific misconception $m$, we search for a program $p$ such that \(\boxed{\forall M, \llbracket p \rrbracket_{\Sigma_{\{m\}}} = \llbracket p \rrbracket_{\Sigma_M} \iff m \in M}.\) In other words, if asked what $p$ outputs, the student should incorrectly respond $\widetilde{r} = \llbracket p \rrbracket_{\Sigma_{\{m\}}}$ if and only if they have the misconception $m$.\footnote{This might not always be possible---for example, some other misconception $m^\prime$ might interact with $m$ in a way that makes it impossible to disentangle the two with a single question. In these cases, we can modify the query to assume that the student does not have misconception $m^\prime$.}

As an example, consider the misconception $m=$ \lstinline{empty-array-is-falsey} (in JavaScript, empty array is actually truthy). A na\"ive diagnostic program to detect $m$ might be $p=$ \lstinline{([] || true)}. Under standard ECMAScript semantics $\Sigma$, this program returns \lstinline{[]} due to the short-circuiting behavior of the \lstinline{||} operator. On the other hand, if a user has misconception $m$ and thinks \lstinline{[]} is falsey, then under $\Sigma_{\{m\}}$ they would expect \lstinline{true}. Hence, $p$ appears to be a reasonable diagnostic: based on their response, we can tell if they have misconception $m$.

However, this is not quite right. Consider a student who has a different misconception $m^\prime=$ \lstinline{||-casts-bool}, under which \lstinline{||} always casts its output to a boolean value. This student would also say $p$ outputs \lstinline{true}, but for a different reason unrelated to the target misconception $m$. Hence, $p$ cannot distinguish between misconceptions $m$ and $m^\prime$, making it a poor diagnostic.

Our system might instead return a program like $p=$ \lstinline{([] ? [] : "abc")}, and furthermore offer a guarantee that the response \lstinline{"abc"} cannot be caused by any other misconceptions (or combinations of misconceptions) known to the system. A list of sample diagnostic programs for each of the misconceptions currently known to \watchatjs---that is, an \emph{automatically-generated concept inventory for JavaScript type coercion}---is given in Tables~\ref{tab:diag-1} and~\ref{tab:diag-2}.

Note that this framework allows for many natural variations: using standard program synthesis techniques, we can search for multiple unique diagnostic programs $p_i$, minimal diagnostic programs, single diagnostic programs that isolate multiple misconceptions (e.g.\ ``power questions'' \citep{reges2008mystery}), and so on with various small modifications to the solver query.

We can also use our system to check if a candidate diagnostic program really does precisely expose a given misconception. We ask the solver for an $M$ that \emph{violates} the earlier condition, i.e. an $M$ such that \(\boxed{\llbracket p \rrbracket_{\Sigma_{\{m\}}} = \llbracket p \rrbracket_{\Sigma_M} \centernot\iff m \in M}.\) If such an $M$ is found by the solver, it can expose one of two issues with $p$:
\begin{enumerate}
\item A set of misconceptions containing $m$ produces a different expected output from $\widetilde{r}$. In this case we get a false negative: a student might have misconception $m$ but not respond $\widetilde{r}$ because some other misconception ``masks'' $m$.
\item A set of misconceptions \emph{not} containing $m$ nevertheless \emph{does} produce $\widetilde{r}$. In this case we get a false positive: a student who does \emph{not} have misconception $m$ might still respond $\widetilde{r}$ because of some other misconception.
\end{enumerate}
Continuing our example from above: if we tell our system that $m$ is \lstinline{empty-array-is-falsy} and $p=$ \lstinline{([] || true)}, then the solver promptly returns $M=\{\texttt{||-casts-bool}\}$ as a counterexample. As explained above, a student with this misconception would indeed expect $p$ to produce \lstinline{true}. This shows that the program $p$ could produce false positives, and is thus a poor diagnostic for $m$.

\section{Instructions and prompts}
\label{sec:prompts}

Below are the complete prompt templates we provided to GPT-4, which mirrored the instructions we provided to human participants in our study. There were only a few minor differences between the GPT-4 prompts and human instructions:
\begin{enumerate}
    \item For the Git scenarios, humans were given the link to the example git repository and prompted to explore it on GitHub or locally by cloning it. GPT-4, however, cannot access the internet when called via API. Hence, we directly provided it with a textual summary of the Git repository contents (the output of \lstinline{ls} in the root directory in each branch).
    \item Humans were told ahead of time that they would complete 8 scenarios (3 JavaScript, 5 Git), whereas GPT-4 was prompted with a single scenario at a time.
    \item Humans were told ``We ask that you do not use AI assistants such as ChatGPT, Copilot, etc. for this study.''
    \item We prefixed GPT-4's system prompt with the following text: ``You are participating in a human participant study. This study is meant for experienced programmers who are fluent in [JavaScript~/~Git, respectively].''
\end{enumerate}

\paragraph{GPT-4 system prompt / human overall instructions}

We used the same system prompt for both Git and JavaScript, substituting \lstinline{SETTING} with the respective domain. 

\begin{quote}
    Imagine that you are working as a front-end web developer at a large company. Because of your technical expertise, you often receive emails from colleagues asking you for help. Because your company is very large, you do not know most of these colleagues — but you do know that their backgrounds range from inexperienced college interns to experienced senior engineers.

    In this study, we will present you with various emails. Each email will be from a different unknown colleague. The colleague will show you some code whose output surprises/confuses them, and then ask you to explain why the code behaves the way it does. We will ask you to respond to each email to the best of your abilities.

    You should directly answer your colleague's question. In particular, you do not need to help your colleague fix the code or improve it. All you should do is explain why it does what it does.

    Sometimes, finding an explanation might be a little tricky. You can in general use any resources you find helpful — including checking documentation, using Google, searching message boards like StackOverflow or Reddit, experimenting with running code yourself, etc.

    You will answer questions about SETTING.
\end{quote}

\paragraph{GPT-4 ``user'' prompt / human per-scenario instructions} Here, we replaced \lstinline{SCENARIO} with the program (JavaScript expression or Git command sequence) and \lstinline{X} and \lstinline{Y} with the true output and expectation, respectively.

\paragraph{JavaScript scenarios}

\begin{quote}
    Imagine you received the following email from a colleague:
    
\begin{verbatim}
Hi there! I have a question about JavaScript.

I just ran the following:

  SCENARIO

Why did it output X? (I expected Y.)

Thanks in advance for your help!
\end{verbatim}
            
            Write a response to your colleague. For the purposes of this study, you can skip pleasantries and get straight to the technical matter.

            Remember: You should directly answer your colleague's question. Do not help your colleague fix or improve the provided code. Only explain the unexpected behavior.
\end{quote}

\paragraph{Git scenarios}

\begin{quote}
    Imagine you received the following email from a colleague:
    
\begin{verbatim}  
Hi there! I have a question about Git.

I just ran the following:

  SCENARIO

Why did I get an error when I ran cat?
  
Thanks in advance for your help!
\end{verbatim}
            
          Write a response to your colleague. For the purposes of this study, you can skip pleasantries and get straight to the technical matter.

          Remember: You should directly answer your colleague's question. Do not help your colleague fix or improve the provided code. Only explain the unexpected behavior.

          Feel free to inspect the repository. The contents of their repository ``https://github.com/watchat-watchat/my-repo.git'' when running ls from each branch is:

            \begin{verbatim}
git checkout 'main'; ls
README.md	output.log

git checkout 'other-work'; ls
otherfile.txt
                
git checkout 'more-work'; ls  
notes.txt
            \end{verbatim}
                
\end{quote}

\begin{table*}
\begin{tabular}{rp{16cm}}
\toprule
\# & Description, expressed as a true statement about JavaScript that corrects the misconception \\ \midrule
1 & For historical reasons, \lstinline[]$null$ has type \lstinline[]$"object"$ (not \lstinline[]$"null"$, as you might expect). \\
2 & Arrays have type \lstinline[]$"object"$ (not \lstinline[]$"array"$, as you might expect). \\
3 & As a special case, \lstinline[]$NaN$ is never equal to anything (even \lstinline[]$NaN$ itself). \\
4 & Empty objects (\lstinline[]${}$ and \lstinline[]$[]$) are truthy. \\
5 & \lstinline$undefined$ is printed as empty string when arrays are cast to string. \\
6 & \lstinline[]$null$ is printed as empty string when arrays are cast to string. \\
7 & \lstinline[]$NaN$ prints as the string \lstinline[]$"NaN"$ (not \lstinline[]$""$, as you might expect). \\
8 & \lstinline[]$null$ casts to the string \lstinline[]$"null"$, not the empty string. \\
9 & undefined casts to the string \lstinline[]$"undefined"$ (not \lstinline[]$""$, as you might expect). \\
10 & \lstinline[]${}$ is \lstinline[]$NaN$ (not 0) when cast to number. \\
11 & JavaScript is 0-indexed, not 1-indexed. \\
12 & \lstinline$undefined$ casts to number as \lstinline[]$NaN$ (not 0, as you might expect). \\
13 & \lstinline[]$null$ casts to number as 0 (not \lstinline[]$NaN$, as you might expect). \\
14 & Array.prototype.sort() casts elements (including numbers) to string and compares them lexicographically. \\
15 & ?? does not treat \lstinline[]$false$ as null-ish. \\
16 & ?? does not treat \lstinline[]$NaN$ as null-ish. \\
17 & The \lstinline[]$+$ operator does not concatenate arrays; instead, it casts them to strings. \\
18 & Short-circuiting boolean operators like \&\& and || return the determining operand (rather than a boolean value). \\
19 & The \lstinline[]$==$ operator, unlike the \lstinline[]$===$ operator, attempts a series of type coercions that can cause unexpected results. \\
20 & When given operands that are neither numbers nor strings, \lstinline[]$+$ tries to cast them to numbers (if possible) or else strings. \\
21 & Objects cast to the string \lstinline[]$"[object Object]"$. \\
22 & When converted to string, arrays don't have the square brackets around them. \\
23 & The empty string by definition casts to 0 (not \lstinline[]$NaN$, as you might expect). \\
24 & The \lstinline[]$+$ operator only attempts to add if both sides are numbers. Otherwise, it casts its operands to string and concatenates. \\
25 & When subscripted, primitive booleans and numbers are implicitly converted to Boolean and Number objects. \\
26 & When one side of an \lstinline[]$==$ is a boolean, JS does not attempt to convert the other side to a boolean as well. Instead, the boolean is converted to a number (0 or 1) and the comparison is tried again. \\
27 & The \lstinline[]$>=$ operator is defined as the negation of \lstinline[]$<$, rather than the disjunction of > and \lstinline[]$==$. \\
28 & If neither operand is a number, then \lstinline[]$<$ compares string representations of the operands lexicographically. \\
29 & The characters \lstinline[]$"["$ and \lstinline[]$"]"$ sort after capital letters but before lowercase letters. \\
30 & The comma character (\lstinline[]$","$) sorts before all letters, numbers, and delimiters. \\
31 & \lstinline[]$==$ and \lstinline[]$===$ compare objects and arrays by reference, not by value. \\
32 & JavaScript casts all indices to string. When indexing arrays and strings, it checks if the indices represent numbers. \\
\bottomrule\end{tabular}
\caption{List of misconceptions currently known to \watchatjs.}\label{tab:misconceptions}
\end{table*}

\begin{table*}
    \centering
    \begin{tabular}{rp{0.9\linewidth}}
\toprule
        \# & Description, expressed as a true statement about Git that corrects the misconception \\ \midrule
        1 & \lstinline$git checkout$ only switches branches. It does not create a new branch if no such branch exists (unless using the \lstinline$-b$ flag).  \\
        2 & \lstinline$git branch$ only creates branches --- it does not move you to the newly-created branch. \\
        3 & \lstinline$HEAD$ is not the same thing as \lstinline$main$, though they are commonly confused. \lstinline$HEAD$ is not a branch; instead, it points to what we have checked out which often points to the most recent commit in the current branch (which may not be \lstinline$main$). \\
        4 & When you started this session, you were on a different branch than you thought you were on. \\
        5 & Git has a notion of a ``staging area'' where you add changes to be recorded as part of the next commit. \\
        6 & \lstinline$git commit$ only commits \emph{staged} changes. \\
        7 & \lstinline$git add$ only \emph{stages} changed files, it does not commit them. \\
        8 & Git will try to carry over uncommitted changes in the staging area when you switch branches. \\
        9 & \lstinline$git rm$ only stages the file's deletion; it does not commit the change. \\
        10 & \lstinline$git rm$ not only stages the file's deletion, but also deletes the file from your disk. (Use \lstinline$--cached$ to avoid this.) \\
        11 & \lstinline$git rm$ can only remove files that are already tracked by Git. You might have meant to use just \lstinline$rm$ (not \lstinline$git rm$). \\
        12 & The shell command \lstinline$rm$ does not stage the file's deletion. To have the deletion tracked, you need to use the Git command \lstinline$git rm$ instead. \\ 
        13 & You can only \lstinline$git mv$ if the source file is already tracked. \\
        14 & The order for \lstinline$git mv$ is source then destination, not destination then source. \\
        15 & \lstinline$git mv$ renames the file both in the repo and your local file system. \\
        16 & \lstinline$git mv$ only stages the file's move; it does not commit the change. \\
        17 & You can't delete a branch (via \lstinline$git branch -d$) without appropriately merging first. \\
        18 & \lstinline$git merge branch-name$ merges branch-name into your current branch, not the other way around. \\
        19 & Running \lstinline$git reset$ with \lstinline$--hard$ both unstages and reverts changes. \\
        20 & \lstinline$git reset$ only unstages changes; to revert the contents of the changed file(s), you need to run \lstinline$--hard$. \\
        21 & \lstinline$git reset$ will not bring back removed files; you need to run  \lstinline$git revert$  to return to the state of a previous commit. \\
        22 & \lstinline$git revert$ will revert the changes from that entire commit; to unstage files, you need to run \lstinline$git revert$. \\
        23 & \lstinline$git diff$ shows the difference between your working directory and your staging area; differences between your staging area and repository history are only shown if you include \lstinline$--staged$. \\
        24 & \lstinline$git add$ does not create a new file. \\
        25 & Git branches are fundamentally different from directories. To switch to a branch, use \lstinline$git checkout$ instead of \lstinline$cd$. \\
        \bottomrule\end{tabular}
    \caption{List of misconceptions currently known to \watchatgit.}
    \label{tab:git_misconceptions}
\end{table*}

\begin{table*}
\centering
\begin{tabular}{rrlll}
\toprule
\# & Time & Synthesized diagnostic program & True output / distractor(s) \\
\midrule
1 & 2 min & \lstinline[]$[typeof(null), {}[[]]]$ & \lstinline[]$["object", undefined]$ \\ & & & \lstinline[]$["null", undefined]$ \\
\midrule2 & 2 min & \lstinline[]$(typeof([]) ?? (false ? false : false))$ & \lstinline[]$"object"$ \\ & & & \lstinline[]$"array"$ \\
\midrule3 & 8 min & \lstinline[]$(NaN === ({} - true))$ & \lstinline[]$false$ \\ & & & \lstinline[]$true$ \\
\midrule4 & 2 min & \lstinline[]$((+undefined) ? {} : ({} ? null : true))$ & \lstinline[]$null$ \\ & & & \lstinline[]$true$ \\
\midrule5 & 4 min & \lstinline[]$([undefined, undefined] + "")$ & \lstinline[]$","$ \\ & & & \lstinline[]$"undefined,undefined"$ \\
& & & \lstinline[]$"[,]"$ \\& & & \lstinline[]$"[undefined,undefined]"$ \\\midrule6 & 5 min & \lstinline[]$("10" + [null, []])$ & \lstinline[]$"10,"$ \\ & & & \lstinline[]$"10null,"$ \\
& & & \lstinline[]$"10[,[]]"$ \\& & & \lstinline[]$"10[null,[]]"$ \\\midrule7 & 10 min & \lstinline[]$(NaN + "10")$ & \lstinline[]$"NaN10"$ \\ & & & \lstinline[]$"10"$ \\
& & & \lstinline[]$NaN$ \\\midrule8 & 3 min & \lstinline[]$("10" + null)$ & \lstinline[]$"10null"$ \\ & & & \lstinline[]$"10"$ \\
\midrule9 & 3 min & \lstinline[]$((false ? undefined : undefined) + "")$ & \lstinline[]$"undefined"$ \\ & & & \lstinline[]$""$ \\
\midrule10 & 2 min & \lstinline[]$[NaN, (+{})]$ & \lstinline[]$[NaN, NaN]$ \\ & & & \lstinline[]$[NaN, 0]$ \\
\midrule11 & 6 min & \lstinline[]$[false, true][1]$ & \lstinline[]$true$ \\ & & & \lstinline[]$false$ \\
\midrule12 & 3 min & \lstinline[]$[({} ?? {}), (+undefined)]$ & \lstinline[]$[{}, NaN]$ \\ & & & \lstinline[]$[{}, 0]$ \\
\midrule13 & 2 min & \lstinline[]$[false, (false - null)]$ & \lstinline[]$[false, 0]$ \\ & & & \lstinline[]$[false, NaN]$ \\
\midrule14 & 29 min & \lstinline[]$[10, 2].sort()$ & \lstinline[]$[2, 10]$ \\ & & & \lstinline[]$[10, 2]$ \\
\midrule15 & 2 min & \lstinline[]$((false ?? true) == false)$ & \lstinline[]$true$ \\ & & & \lstinline[]$false$ \\
\midrule16 & 3 min & \lstinline[]$(({} - undefined) ?? (!true))$ & \lstinline[]$NaN$ \\ & & & \lstinline[]$false$ \\
\midrule
\end{tabular}
\caption{A concept inventory (diagnostic ``quiz'') automatically synthesized by \watchatjs. For each misconception known to \watchatjs, we synthesize a multiple-choice question where the first choice is correct, and the second choice would be chosen \emph{if and only if} the student has the respective misconception. In some cases, our system can generate other distractors as well. (Misconceptions numbered as in Table~\ref{tab:misconceptions}. Continued below in Table~\ref{tab:diag-2}.)}\label{tab:diag-1}
\end{table*}

\begin{table*}[t]
\centering
\begin{tabular}{rrlll}
\toprule
\# & Time & Synthesized diagnostic program & True output / distractor(s) \\
\midrule
17 & 2 min & \lstinline[]$[(!true), ([] + [])]$ & \lstinline[]$[false, ""]$ \\ & & & \lstinline[]$[false, []]$ \\
& & & \lstinline[]$(error)$ \\& & & \lstinline[]$[false, "[][]"]$ \\\midrule18 & 2 min & \lstinline[]$((+true) || NaN)$ & \lstinline[]$1$ \\ & & & \lstinline[]$true$ \\
\midrule19 & 3 min & \lstinline[]$[(undefined == null), (+true)]$ & \lstinline[]$[true, 1]$ \\ & & & \lstinline[]$[false, 1]$ \\
\midrule20 & 1 min & \lstinline[]$((!true) ? ({} ? {} : false) : (undefined + null))$ & \lstinline[]$NaN$ \\ & & & \lstinline[]$(error)$ \\
& & & \lstinline[]$0$ \\\midrule21 & 6 min & \lstinline[]$("0" + {})$ & \lstinline[]$"0[object Object]"$ \\ & & & \lstinline[]$"0{}"$ \\
\midrule22 & 7 min & \lstinline[]$("" >= [])$ & \lstinline[]$true$ \\ & & & \lstinline[]$false$ \\
\midrule23 & 2 min & \lstinline[]$("" - "")$ & \lstinline[]$0$ \\ & & & \lstinline[]$NaN$ \\
\midrule24 & 3 min & \lstinline[]$([undefined, {}] + ([] - true))$ & \lstinline[]$",[object Object]-1"$ \\ & & & \lstinline[]$NaN$ \\
& & & \lstinline[]$",[object Object]NaN"$ \\& & & \lstinline[]$"[,[object Object]]NaN"$ \\& & & \lstinline[]$",[object Object]"$ \\& & & \lstinline[]$",{}-1"$ \\& & & \lstinline[]$"undefined,[object Object]-1"$ \\\midrule25 & 2 min & \lstinline[]$(false[[]] === {})$ & \lstinline[]$false$ \\ & & & \lstinline[]$(error)$ \\
\midrule26 & 5 min & \lstinline[]$((undefined == false) || (undefined == {}))$ & \lstinline[]$false$ \\ & & & \lstinline[]$true$ \\
\midrule27 & 3 min & \lstinline[]$("," >= (+false))$ & \lstinline[]$true$ \\ & & & \lstinline[]$false$ \\
\midrule28 & 4 min & \lstinline[]$("2" >= "10")$ & \lstinline[]$true$ \\ & & & \lstinline[]$false$ \\
\midrule29 & 33 min & \lstinline[]$("[" < typeof(true))$ & \lstinline[]$true$ \\ & & & \lstinline[]$false$ \\
\midrule30 & 19 min & \lstinline[]$("," >= [{}, []])$ & \lstinline[]$false$ \\ & & & \lstinline[]$true$ \\
\midrule31 & 2 min & \lstinline[]$("" || ([] == []))$ & \lstinline[]$false$ \\ & & & \lstinline[]$true$ \\
\midrule32 & 3 min & \lstinline[]$[[], {}]["1"]$ & \lstinline[]${}$ \\ & & & \lstinline[]$undefined$ \\
& & & \lstinline[]$[]$ \\\midrule
\end{tabular}
\caption{(Continued from above in Table~\ref{tab:diag-1}.)}\label{tab:diag-2}
\end{table*}

\end{document}